\documentclass[preprint, authoryear, 12pt]{elsarticle}
\usepackage{lineno}  %hyperref
\modulolinenumbers[1]
\usepackage{setspace}

\journal{*****}

\usepackage{natbib}
\usepackage{amsmath}
\usepackage[linesnumbered,ruled,algo2e]{algorithm2e}
\usepackage{mathabx}
\usepackage{booktabs}
\usepackage{balance}
\usepackage{color}
\usepackage{soul}
\usepackage{subcaption}
\usepackage[margin=1in]{geometry}
\usepackage{url}
\usepackage[most]{tcolorbox}
\definecolor{mybg}{RGB}{192,255,220}
\newcommand\fscore{\mathop{\mbox{$F$-$\mathit{score}$}}}

\makeatletter
\def\ps@pprintTitle{%
	\let\@oddhead\@empty
	\let\@evenhead\@empty
	\def\@oddfoot{\centerline{\thepage}}%
	\let\@evenfoot\@oddfoot}
\newcommand*{\centerfloat}{
  \parindent \z@
  \leftskip \z@ \@plus 1fil \@minus \textwidth
  \rightskip\leftskip
  \parfillskip \z@skip}
\makeatother

\usepackage{changes}
\usepackage{comment}
\usepackage{todonotes}
\usepackage{listings}
\usepackage{courier}
\usepackage{float}

\usepackage{tikz}
\def\checkmark{\tikz\fill[scale=0.4](0,.35) -- (.25,0) -- (1,.7) -- (.25,.15) -- cycle;}

\newcommand\blfootnote[1]{%
	\begingroup
	\renewcommand\thefootnote{}\footnote{#1}%
	\addtocounter{footnote}{-1}%
	\endgroup
}

\begin{document}

\begin{frontmatter}

\title{Enhancing spatial and textual analysis with EUPEG: an extensible and unified platform for evaluating geoparsers}

\author{Jimin Wang and Yingjie Hu}
% \ead{jiminwan@buffalo.edu}

%\author{Yingjie Hu}
% ead{yhu42@buffalo.edu}

\address{GeoAI Lab, Department of Geography, University at Buffalo, NY 14260, USA}

\begin{abstract}
A rich amount of geographic information exists in unstructured texts, such as Web pages, social media posts,  housing advertisements,  and historical archives. Geoparsers are useful tools that extract structured geographic information from unstructured texts, thereby enabling spatial analysis on textual data. While a number of geoparsers were developed, they were tested on different datasets  using different  metrics. Consequently, it is difficult to compare existing geoparsers or to compare a new geoparser with existing ones. In recent years, researchers created open and annotated corpora for testing geoparsers. While these corpora are extremely valuable, much effort is still needed for a researcher to prepare these datasets and deploy geoparsers for comparative experiments. This paper presents EUPEG: an Extensible and Unified Platform for Evaluating Geoparsers. EUPEG is an open source and Web based benchmarking platform which hosts a majority of open corpora, geoparsers, and performance metrics reported in the literature. It enables direct comparison of the hosted geoparsers, and a new geoparser can be connected to EUPEG and compared with other geoparsers. The main objective of EUPEG is to reduce the time and effort that researchers have to spend in preparing datasets and baselines, thereby increasing the   efficiency and effectiveness of comparative experiments.

\end{abstract} 


\end{frontmatter}

% \linenumbers
%\openup 0.6em

\blfootnote{\noindent \\Wang, J. and Hu, Y. (2019): Enhancing spatial and textual analysis with EUPEG: an extensible and unified platform for evaluating geoparsers, \textit{Transactions in GIS}, accepted. \\ DOI: \url{https://www.doi.org/10.1111/tgis.12579} \\
Contact info: Jimin Wang (jiminwan@buffalo.edu); Yingjie Hu (yhu42@buffalo.edu)}

\section{Introduction} 

Many studies and applications nowadays need an integration of spatial and textual analysis. In disaster response, it is often necessary to recognize, geo-locate, and analyze the place names mentioned in short text messages or social media posts in order to understand who needs help and where \citep{maceachren2011senseplace2,gelernter2013algorithm,lan2014spatio,pezanowski2018senseplace3}. Studying place relations and interactions in virtual or cognitive spaces usually involves extracting place names from texts, such as Wikipedia pages or news articles, and analyzing their co-occurrences with spatial distances \citep{hecht2009terabytes,liu2014analyzing,geiss2015wikipedia,salvini2016spatialization,hu2017extracting}.  To develop place-based GIS, researchers may need to examine human experiences encoded in texts, such as in travel blogs, and how these human experiences are related to spatial locations \citep{adams2013inferring,ballatore2015extracting,gao2017data}.   In addition, there exists various  geographic knowledge in Web pages \citep{jones2008modelling}, housing advertisements \citep{mckenzie2018identifying}, business documents \citep{faulconbridge2008new}, historical archives \citep{grossner2016place}, and other types of texts. %Discovering knowledge from these data sources often needs an integration of spatial and textual analysis.

%For example, analyzing historical archives often requires identifying old place names from texts and then performing spatial analysis on the geocoded place names \citep{grossner2016place,barbaresi2017towards}. Studying emerging events based on news articles or social media posts also frequently needs to recognize and map place names mentioned in their textual content . There also exists useful geographic information in  \textit{Geoparsing} is a critical step for extracting structured geographic information from unstructured texts, and therefore enables spatial analysis on the initially non-spatial textual data. 

\mbox{\textit{Geoparsing} is a critical process for extracting  spatial information from textual data.}  It is recognized as an important research topic in the broader field of geographic information retrieval (GIR) \citep{doi:10.1080/13658810701626343,purves2018geographic}. A  geoparsing system is called a \textit{geoparser} which takes unstructured textual data as the input and outputs a set of recognized place names and their  spatial footprints \citep{freire2011metadata,leidner2008toponym}.

While a number of geoparsers have already been developed, it is difficult to directly compare their performances. Examples of the developed geoparsers include MetaCarta \citep{frank2006spatially}, GeoTxt \citep{karimzadeh2013geotxt}, the Edinburgh Geoparser \citep{alex2015adapting}, TopoCluster \citep{delozier2015gazetteer}, and CamCoder \citep{gritta2018melbourne}. Two factors make the direct comparison difficult. First,  many  geoparsers were tested on project-specific datasets that are not shared publicly. Besides the additional effort  for making data ready for sharing, there also exist policy restrictions (e.g., Twitter forbids  sharing the content of tweets) and privacy concerns that prevent researchers from  sharing their data publicly. As a result, geoparsers cannot be fairly compared since the same geoparser can have very different performances  depending on the testing datasets \citep{leidner2006evaluation,ju2016things}. Second, different performance metrics were usually used for evaluating geoparsers.  Some researchers used the metrics of \textit{precision}, \textit{recall}, and \textit{F-score} adopted from the field of information retrieval, while some others used metrics based on spatial distances, such as \textit{mean} or \textit{median error distance}. %Even when the same metrics were used, it can be unclear whether researchers used the same rules for determining a match between a geoparsed toponym and the ground-truth annotation (e.g., exact or inexact matching). %, e.g., there are both micro F-score and macro F-score \citep{cornolti2013framework}. 
Due to both factors, we cannot  compare geoparsers by  juxtaposing the performance numbers reported in  their papers.

To effectively compare existing geoparsers or to compare a new geoparser with existing baselines, one would ideally find and deploy the geoparsers in the literature and use the same datasets and metrics to test their performances. However, such a process is time-consuming and labor-intensive. First, one needs to find openly shared and annotated datasets. While the community has already made great efforts to share datasets,  such as  the Local-Global Lexicon (LGL) corpus \citep{lieberman2010geotagging}, WikToR \citep{gritta2018s}, and GeoCorpora \citep{doi:10.1080/13658816.2017.1368523}, researchers still need to spend much time preparing these datasets for experiments. For example, GeoCorpora is a valuable dataset containing human annotated tweets.  Due to Twitter's policy restriction, GeoCorpora contains only the IDs of tweets rather than their full content. To use GeoCorpora, one needs to apply for a developer account for using Twitter's Application Programming Interface (API) %(which involves writing an application statement, waiting for reviews, and receiving a decision regarding approval or rejection) 
and write a program to \textit{rehydrate} the dataset. Even for the datasets that are more readily available, they are often in different formats and need to be harmonized into the same format before an experiment.  Second, it takes a considerable amount of time to find, deploy, and re-run existing geoparsers. %Researchers often need to find the source code (or compiled binary versions) of the geoparsers, download the code, read the installation instructions and related documents, deploy the system, and address any configuration issues. 
This process can take even longer when no direct download link is provided for a geoparser or when there is a lack of deployment instructions. Third, a set of  performance metrics need to be implemented to compare geoparsers. Although implementing these metrics may not be difficult, one needs to harmonize the heterogeneous output formats of different geoparsers and compare their  outputs with ground truth annotations. In sum, conducting an effective comparative experiment of geoparsers costs a lot of time and human resources. While those costs are probably fine for a single research group,   the entire community can lose a lot of  time if every individual research group  has to prepare datasets, geoparsers, and metrics in order to run an experiment.

This paper presents EUPEG: an Extensible and Unified Platform for Evaluating Geoparsers. EUPEG is  designed as an open source and Web based platform. It hosts a majority of the open corpora, geoparsers, and performance metrics reported in the literature. One can directly compare these hosted geoparsers on the same datasets using the same metrics, or can connect a new geoparser to EUPEG and compare it with the existing ones. The value of EUPEG can be seen from the perspectives of both geoparser users and  researchers. For a user who would like to find a suitable geoparser to process a corpus, EUPEG offers a comprehensive view on the advantages and limitations of different geoparsers (e.g., some may have higher precision while some others may have higher recall) and their performances on different types of corpora (e.g., short messages or long articles). For  researchers who would like to develop a new geoparser, they can focus on inventing new methods rather than  preparing datasets and  baselines. In addition, EUPEG automatically archives the results and configurations of experiments, such as the date and time of an experiment, the selected datasets, used geoparsers, and performance metrics. Researchers can share experiment results with their colleagues or even the general public more easily via an experiment ID. The contributions of this paper are as follows:

\begin{itemize}
	\item We propose and develop a benchmarking platform, EUPEG, for effective and efficient comparison of geoparsers. EUPEG currently hosts eight annotated geographic corpora, nine geoparsers, and eight performance metrics. New geoparsers and datasets can also be connected to it. Experiment results and configurations are recorded and can be shared via experiment IDs. A demo of EUPEG can be accessed at:  \url{https://geoai.geog.buffalo.edu/EUPEG}.
	 
   \item  We provide a systematic review  on the  geoparsing resources  hosted on EUPEG.  The  corpora  are in four different text genres, ranging from news articles to social media posts; the geoparsers are developed using different methods, such as heuristics and machine learning; and the performance metrics are from information retrieval or based on spatial distances. EUPEG serves as a one-stop platform that unifies the heterogeneous datasets, geoparsers, and performance metrics.

	\item We share the source code of EUPEG on GitHub, along with the hosted resources under permitted licenses (e.g., GNU General Public License). The code repository can be accessed at: \url{https://github.com/geoai-lab/EUPEG}. The shared source code enables researchers to run EUPEG on a local computer, or to add more datasets and geoparsers at the source-code level. One can also extend EUPEG with new features, such as new performance metrics  suitable for a project.

	% or to extend EUPEG with new geoparsing resources and to add new features for enhancing the benchmarking platform itself. The source code can also facilitate the development of benchmarking platforms for other  research problems in geography, such as land use and land cover (LULC) classification, where different  solutions are proposed for the same problem. }
	
%	\item  \hl{ EUPEG along with this paper, can be viewed as a detailed investigation of currently available corpora, geoparers, and evaluation metrics. Based on the review, we provides a scheme to fit across the difference of all these resources and methods. The idea and design behind EUPEG can be generalized to other research problems in geography, such as land use and land cover (LULC) classification, in which multiple different solutions are proposed for the same problem.}
\end{itemize}

While EUPEG is designed for geoparsing, the idea of developing benchmarking platforms can be extended to other  research topics in geography, such as land use and land cover (LULC) classification, where different  solutions are developed for addressing the same problem. This paper is a major extension of our previous short paper \citep{Hu:2018:ETE:3281354.3281357}. The remainder of this paper is organized as follows. Section 2 reviews related work on geoparsing, corpora building, and benchmarking platforms. Section 3 presents the design details of EUPEG, including the overall architecture and the hosted datasets, geoparsers, and performance metrics. %We also discuss the schemas and formats of new datasets and geoparsers  to be connected to EUPEG. 
Section 4 demonstrates the implemented EUPEG and provides an analytical evaluation on the approximate time that can be saved by EUPEG for comparative experiments. Finally, Section 5 summarizes this work and discusses future directions.

\section{Related Work}
In this section, we provide a review on related studies. We start by introducing the background knowledge of geoparsing and major geoparsers developed so far, and  continue to discuss the efforts made by the  community to create and share open and annotated corpora.   We then discuss the  recent movement towards developing benchmarking platforms for effective and efficient comparisons of different solutions to the same  problems.

\subsection{Geoparsing and geoparsers}
\begin{figure}[h]
 	\centering
	\includegraphics[width=0.9\textwidth]{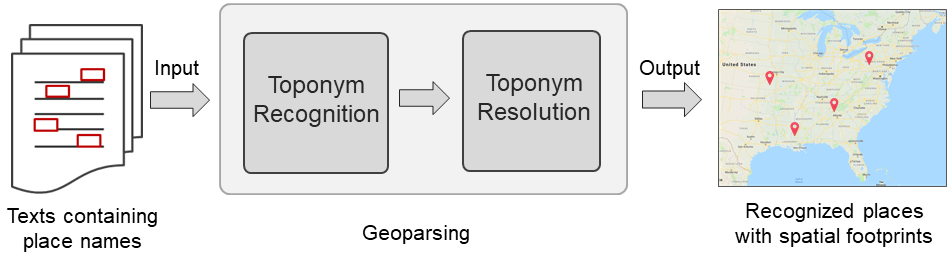}
	\caption{The input and output of geoparsing and its two main steps.}\label{geoparsing}
	%	\vspace*{-0.5cm}
\end{figure}

Geoparsing is a research topic often studied in GIR  \citep{doi:10.1080/13658810701626343,purves2018geographic}. The goal of geoparsing is to recognize place names mentioned in texts and resolve them to the corresponding place instances and location coordinates  \citep{freire2011metadata,barbaresi2017towards,gritta2018s}. Geoparsing is typically performed in two consecutive steps: toponym recognition and toponym resolution. The first step recognizes place names from  texts without identifying the particular place instance referred by a name, while the second step aims to resolve any ambiguity of the place name and locates it to the right spatial footprint. Figure \ref{geoparsing} illustrates the input and output of geoparsing and its two steps.

Many methods have been proposed for these two steps. For toponym recognition, early research used gazetteer-based entry matching and grammatical rules \citep{woodruff1994gipsy,purves2018geographic}, while more recent approaches employed machine learning and natural language processing techniques. Particularly, the Stanford Named Entity Recognition (NER) tool was used in many studies for  toponym recognition \citep{lieberman2010geotagging,gelernter2011geo,karimzadeh2013geotxt,delozier2015gazetteer}. For toponym resolution,  various heuristics were developed to resolve place name ambiguity \citep{amitay2004web,leidner2008toponym}. A simple method is to resolve a place name to the place instance that has the highest population or the largest total geographic area  \citep{li2002location,ladra2008toponym}. Machine learning models were also developed for toponym resolution by exploiting various features, such as toponym co-occurrences \citep{overell2008using},  words in the local context \citep{speriosu2013text},  distances among toponyms \citep{santos2015using}, topics of the local context \citep{ju2016things}, and a combination of multiple features \citep{nesi2016geographical,gritta2018melbourne}.

A number of geoparsers were developed which can function as end-to-end systems for completing both steps. \textit{GeoTxt}, developed by  \cite{karimzadeh2013geotxt}, is a Web-based geoparser that leverages Stanford NER  and two other NER tools  for toponym recognition and uses GeoNames and a set of heuristic rules for toponym resolution. TopoCluster, by \cite{delozier2015gazetteer},  can perform geoparsing without using a gazetteer. It uses Stanford NER to recognize toponyms from texts and then resolves toponyms based on the \textit{geographic profiles} of words in the surrounding context. Cartographic Location And Vicinity INdexer (CLAVIN) is an open source geoparser that employs Apache OpenNLP  for toponym recognition and utilizes a gazetteer and fuzzy search for toponym resolution. The Edinburgh Geoparser is a geoparsing system developed by the Language Technology Group at Edinburgh University \citep{alex2015adapting}. It uses their in-house software tool, called LT-TTT2, for toponym recognition, and the toponym resolution step is based on a gazetteer such as GeoNames. CamCoder is a deep learning based geoparser developed by \cite{gritta2018melbourne}, which integrates convolutional neural networks, word embeddings, and the geographic vector representations of place names. There also exist commercial geoparsers, such as Geoparser.io (\url{https://geoparser.io}), which often charge a fee. Some commercial geoparsers, such as Yahoo! PlaceSpotter (\url{https://developer.yahoo.com/boss/geo/docs/PM_KeyConcepts.html}), provide relatively permissive rate limitations for free requests (e.g., 2,000 calls per hour).

\subsection{Efforts in sharing open and annotated corpora}
While many geoparsers exist, it is difficult to directly compare them due to a lack of open and annotated corpora. In recent years, researchers made great efforts to address this issue. %In an early work, \cite{leidner2006evaluation} developed the TR-CoNLL corpus which contains 946 annotated news articles from Reuters. %\cite{leidner2006evaluation} also suggested that ideally different text genres should  be included in the test corpora to provide more comprehensive evaluations on geoparsers.  
\cite{lieberman2010geotagging} shared a human annotated dataset called Local-Global Lexicon (LGL) containing 588 news articles published by local newspapers from highly ambiguous places. \cite{hu2014improving} contributed an automatically annotated corpus containing short sentences retrieved from the home pages of cities with ambiguous names such as \textit{Washington}. \cite{ju2016things} shared a corpus of short sentences from various Web pages, which was automatically collected and annotated using a script based on the Microsoft Bing Search API. \cite{gritta2018s} contributed WikToR which is a corpus of Wikipedia articles with ambiguous names, such as \textit{Lima, Peru} and \textit{Lima, Oklahoma}, automatically annotated by a Python script.   \cite{doi:10.1080/13658816.2017.1368523} contributed GeoCopora which is a dataset of tweets manually annotated using a  hybrid approach with both users on Amazon's Mechanical Turk and researchers in the domain of geography. \cite{gritta2018melbourne} and \cite{gritta2018pragmatic} shared two human annotated corpora, GeoVirus and GeoWebNews, which contain 229 and 200 news articles respectively. TR-News is another news article corpus  which contains 118  articles manually annotated by \cite{kamalloo2018coherent}.  In addition to contemporary corpora, some historical datasets were also made available, such as \textit{War Of The Rebellion} (WOTR) by \cite{delozier2016creating}. \cite{leidner2006evaluation} developed the TR-CoNLL corpus which contains 946 annotated news articles from Reuters; however, it is not publicly available to the best of our knowledge. The ACE 2005 English SpatialML is an annotated news corpus shared on the Linguistic Data Consortium  \citep{mani2008ace}, but it charges a fee (\$1,000) for non-members. While these  annotated corpora greatly facilitate the development and testing of geoparsers, finding, downloading, and preparing these corpora require considerable amounts of time and effort. %For example,  datasets are usually annotated using different formats, and one has to harmonize such heterogeneous formats before conducting an experiment. 

\subsection{Benchmarking platforms}

The importance and necessity of evaluating geoparsers in a systematic  manner have already been recognized by the research community \citep{monteiro2016survey,melo2017automated,richter2017heidelplace,gritta2018s,doi:10.1080/13658816.2017.1368523}. \cite{melo2017automated} argued that the fact that one geoparser performed worse than another geoparser on one particular dataset did not mean that this geoparser would still perform worse than the other if a different dataset were used. %obtaining bad result on one specific dataset does not stand for that it also performances worse on other datasets.} 
\cite{gritta2018s} compared the performances of five geoparsers on two corpora using a set of standard performance measures, such as precision, recall, F-score, and median error distance. In their more recent work, the authors further proposed a pragmatic guide to geoparsing evaluation \citep{gritta2018pragmatic}. In addition to publications, \cite{gritta2018s,gritta2018pragmatic}  also released their source code and the annotated corpora which greatly facilitated the reproduction of their experiments. EUPEG is built on the  foundational work of \citep{gritta2018s,gritta2018pragmatic}, but extends their work in three aspects. First, EUPEG provides a benchmarking platform which offers datasets and baseline geoparsers that are ready for use. While \cite{gritta2018s,gritta2018pragmatic} have shared the source code of comparing five geoparsers, a lot of effort is still needed to understand, deploy, and run these geoparsers. Some geoparsers do not function on certain operating systems (OS) (e.g., the Edinburgh Geoparser is not supported on Windows) or require extra database configurations (e.g., TopoCluster requires PostgreSQL and PostGIS), which can add additional requirements on their deployments. EUPEG directly hosts  these geoparsers, along with annotated corpora and performance metrics. Researchers can directly run experiments on EUPEG, and can connect their own geoparsers and datasets to the platform. Second, EUPEG extends the corpora and geoparsers from \cite{gritta2018s,gritta2018pragmatic}. We provide eight annotated corpora in four different text genres, which include news articles, Wikipedia articles, social media posts, and Web pages. We provide nine geoparsing methods which include not only specialized geoparsers (e.g., GeoTxt and CLAVIN) but also a number of geoparsing systems extended from general NER tools, such as DBpedia Spotlight, Stanford NER, and spaCy NER. Third, EUPEG offers the  capability of archiving research experiments. Each  experiment is assigned a unique  ID that allows researchers to share first-hand research outcomes and to search the results of previous experiments.

The demand for benchmarking platforms is also witnessed in other research fields beyond  geography. \cite{cornolti2013framework} developed a framework for systematically evaluating a number of named entity annotators, such as AIDA, Illinois Wikifier, and DBpedia Spotlight, on the same datasets. Building on the work of \cite{cornolti2013framework}, \cite{usbeck2015gerbil} developed GERBIL which is a platform for   agile, fine-grained, and uniform evaluations of named entity annotation tools. The practice of comparing different methods on the same datasets was also seen in  computer science conferences, such as  the Message Understanding Conference (MUC) \citep{sundheim1993tipster}, the Conference on Computational Natural Language Learning (CoNLL) \citep{tjong2003introduction}, and the Making Sense of Microposts workshop series (MSM) \citep{cano2014making}. Such a practice is especially effective when multiple solutions exist for the same research problem, and can reveal the advantages and limitations of different solutions. Sharing datasets for comparing  methods can fuel the advancement in a particular research area as well. For example, the availability of the ImageNet dataset was a critical boost to  the remarkable development of deep learning  in computer vision \citep{deng2009imagenet}. To the best of our knowledge, EUPEG is the first benchmarking platform for the research problem of geoparsing. %The idea of developing benchmarking platforms is not limited to geoparsing but can be generalized to other problems in geographic research as well, such as LULC classification based on remote sensing images or event detection based on social media data. 

%Starting in 1993, the Message Understanding Conference (MUC) introduced a first systematic comparison of information extraction approaches [42]. Ten years later, the Conference on Computational Natural Language Learning (CoNLL) started to offer a shared task on named entity recognition and published the CoNLL corpus [43]. In addition, the Automatic Content Extraction (ACE) challenge [10], organized by NIST, evaluated several approaches but was discontinued in 2008. Since 2009, the text analytics conference hosts the workshop on knowledge base population (TAC-KBP) [22] where mainly linguistic-based approaches are published. The Senseval challenge, originally concerned with classical NLP disciplines, has widened its focus in 2007 and changed its name to SemEval to account for the recently recognized impact of semantic technologies [19]. The Making Sense of Microposts workshop series (MSM) established in 2013 an entity recognition and in 2014 an entity linking challenge thereby focusing on tweets and microposts [37]. In 2014, Carmel et al. [5] introduced one of the first Web-based evaluation systems for NER and NED and the centerpiece. 

%The need to systematically evaluate geoparsers based on the same datasets and same performance measures was recognized by researchers 

%ImageNet

%- Computer science competitions make major progress

%\vspace*{-0.4cm}
\section{EUPEG}

\subsection{Overall architecture}

EUPEG is designed as a Web based and open source benchmarking platform. It provides two main functions:
\begin{itemize}
	\item It enables effective and efficient comparison of different geoparsers on the same datasets using the same performance metrics. 
	\item It facilitates the sharing of experiments by archiving evaluation results and configurations and supporting the search of previous experiments.   
\end{itemize} 

\begin{figure}[h]
 	\includegraphics[width=\textwidth]{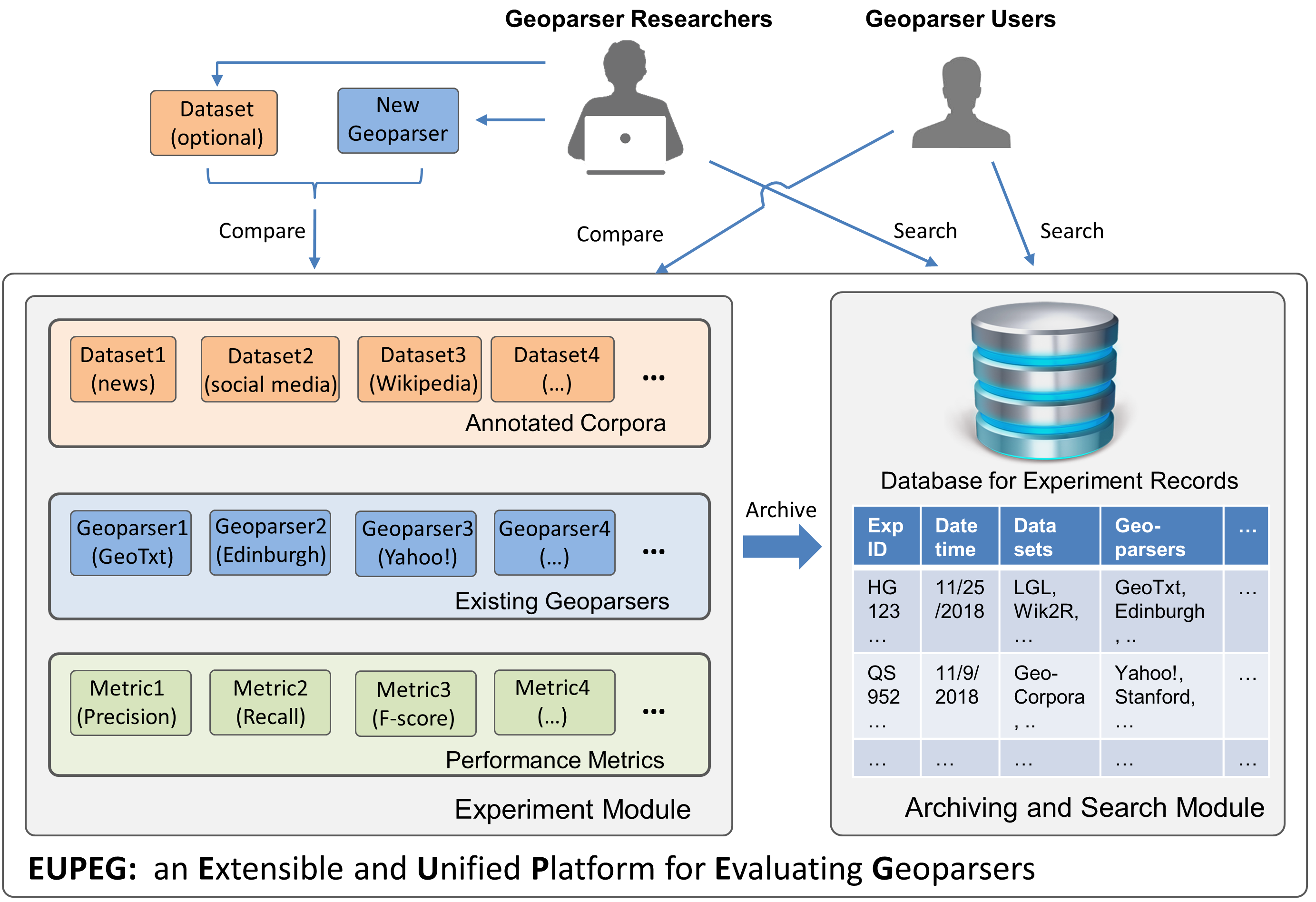}
 	\caption{The overall architecture of EUPEG.}\label{overall}
 	%	\vspace*{-0.5cm}
 \end{figure}

The overall architecture of EUPEG is shown in Figure \ref{overall}.
Two major modules are designed. The \textit{Experiment Module} hosts a majority of openly available resources including annotated corpora, existing geoparsers, and  performance metrics. The \textit{Archiving and Search Module} records experiment results and supports the search of previous experiments. For geoparser researchers who would like to develop new geoparsers, they can  connect a new geoparser to EUPEG and compare it with  others    on the hosted datasets using the same performance metrics. Researchers can also upload one or multiple customized datasets and compare the new geoparser with existing baselines on these customized datasets. For users who would like to find a suitable geoparser for processing a corpus, they can  compare existing geoparsers directly on EUPEG to see their advantages and limitations. The experiment configurations (e.g., the used datasets, geoparsers, and metrics) and results are recorded in a database in the \textit{Archiving and Search Module}. One can search previous experiments based on their experiment IDs  automatically generated by EUPEG. In the following, we present  details about the datasets, geoparsers, and metrics.

\subsection{Datasets}

EUPEG hosts a majority of annotated geographic corpora reported in the literature. Two criteria are used for selecting these datasets. First, they have to be formally described by  existing papers; second, the datasets should be openly available without a fee (e.g., a dataset shared on GitHub under the MIT license). In addition, we focus on \textit{geographic} corpora, namely those  with  toponyms and spatial footprints annotated. There exist general NER corpora whose annotations contain other types of entities (e.g., persons and concepts) and  do not provide spatial footprints. Those general NER corpora are not included.

 %There  exist datasets for general named entity recognition purposes, which contain other types of entities, such as persons, facilities, and vehicles, in addition to locations. While these general NER datasets could be converted to geographic corpora by retaining the location annotations only and adding geographic coordinates. Doing so is beyond the scope of this work and can introduce extra noise and other uncertain factors. Therefore, we do not include general NER datasets. 

The corpora hosted on EUPEG are in four different text genres: news articles, Wikipedia articles, social media posts, and Web pages. Having multiple genres rather than a single text type can provide a more comprehensive evaluation on the performance of a geoparser \citep{leidner2006evaluation}.  These datasets are described as follows.

\textbf{News articles}.  %Annotated news articles were provided by multiple previous studies. 
Four news article corpora are hosted on EUPEG:
\begin{itemize}
	\item \textit{LGL}. LGL was developed by  \cite{lieberman2010geotagging}, which contains 588 human-annotated news articles published by 78 local newspapers from highly ambiguous places, such as \textit{Paris News} (Texas), \textit{Paris Post-Intelligencer} (Tennessee), and \textit{Paris Beacon-News} (Illinois). %LGL considers demonyms (e.g. Canadian) as toponyms and annotates them to point coordinates (e.g., the center point of Canada).
	
	\item \textit{GeoVirus}. This is a human-annotated dataset  shared by \cite{gritta2018melbourne}. It contains 229 news articles from WikiNews during 08/2017 - 09/2017. These news articles cover global disease outbreaks and epidemics, and were collected using keywords, such as \textit{``Ebola"} and \textit{``AIDS"}. %Building names, point-of-interest (POI) names, street names, and river names were not considered as toponyms and are not annotated.

	\item \textit{TR-News}. This dataset was contributed by \cite{kamalloo2018coherent}, which contains 118 human-annotated news articles from various global and local news sources. The authors deliberately included less dominant place instances, such as \textit{Edmonton, England} and \textit{Edmonton, Australia}, while also keeping articles from general global news sources, such as BBC and Reuters.  
	
	\item \textit{GeoWebNews}. This dataset was shared by \cite{gritta2018pragmatic} which comprises 200 human-annotated news articles from 200 globally distributed news sites collected from April 1st to 8th in 2018. The authors randomly selected one article from each domain until they reached 200 news articles for creating this dataset. 
\end{itemize}

\textbf{Wikipedia articles.} One Wikipedia article corpus is hosted on EUPEG.
\begin{itemize}
	\item \textit{WikToR}. This corpus was provided by  \cite{gritta2018s}. It was automatically generated by a script using those Wikipedia articles about places. It contains 5,000 articles with ambiguous names, such as \textit{Lima, Peru}, \textit{Lima, Ohio}, and \textit{Lima, Oklahoma}. One limitation of WikToR is that it does not annotate all toponyms in the texts but only those that are the description target of a Wikipedia article. As a result, some performance metrics, such as precision, recall, and F-score, cannot be used for quantifying the performance of geoparsers based on WikToR. 
\end{itemize}

\textbf{Social media posts.} EUPEG hosts one social media dataset, GeoCorpora, contributed by \cite{doi:10.1080/13658816.2017.1368523}. 
\begin{itemize}
	\item \textit{GeoCorpora.} This is a tweet corpus that contains 1,639 human-annotated tweet posts. These posts were first annotated using a crowdsourcing approach by  workers on Amazon's Mechanical Turk and then further annotated (with disagreements resolved) by researchers in geography. It is worth noting that the original paper reported 2,122 tweets with toponyms annotated. Due to the data sharing  restriction of Twitter, \cite{doi:10.1080/13658816.2017.1368523} could not share the full content of tweets but only tweet IDs. When \textit{rehydrating} tweets, we were able to recover only 1,639 tweets, since some tweets were deleted by their authors. The number of tweets that can be recovered will only decrease as the time goes by. Thus, we believe that another value of EUPEG is its capability of preserving valuable contributions from previous research by sidestepping some policy restrictions (in this case, EUPEG does not provide any direct download of the tweets). %and thus does not violate the data policy of Twitter).
\end{itemize}

\textbf{Web pages.} %Geographically annotating and indexing Web pages are among the major motivations of geoparsing \citep{amitay2004web}. 
Two Web page corpora are hosted on EUPEG.

\begin{itemize}
	\item \textit{Hu2014}. This is a small corpus contributed by \cite{hu2014improving}, which was automatically constructed by a script. The authors focused on two highly ambiguous US place names, \textit{Washington} and \textit{Greenville}, and retrieved textual descriptions from the websites of  related cities. The texts in these Web pages were then divided into shorter sentences. Overall, this dataset contains 134 entries. Not all toponyms in the sentences are annotated, and therefore precision, recall, and F-score cannot be applied to evaluating the geoparsing results based on this dataset (similar to \textit{WikToR}). 
	
	\item \textit{Ju2016}. This is another automatically constructed corpus. It was contributed by \cite{ju2016things} who made use of a list of highly ambiguous US place names on Wikipedia, and then used Microsoft Bing Search API to retrieve sentences from various Web pages (Wikipedia articles were removed from these Web pages) that contain the searched place names. This corpus contains 5,441 entries. Similar to \textit{WikToR} and \textit{Hu2014}, this dataset does not annotate all toponyms  and cannot use the performance metrics of precision, recall, and F-score.
\end{itemize}

In total, EUPEG hosts eight geographic corpora in four different text genres. Table \ref{dataset_table} summarizes the attributes of these datasets. 
\begin{table}[h]
	\caption{A summary of the open and annotated corpora hosted on EUPEG.}
	\begin{tabular}{rccccc}
		\hline
		Dataset    & Genre        & \begin{tabular}[c]{@{}c@{}}Text \\ Date\end{tabular} & \begin{tabular}[c]{@{}c@{}}Entry\\ Count\end{tabular} & \begin{tabular}[c]{@{}c@{}}Average Words\\ per Entry\end{tabular} & \begin{tabular}[c]{@{}c@{}}Average Toponym\\ per Entry\end{tabular} \\ \hline
		LGL        & News         & 03/2009                                              & 588                                                   & 315                                                               & 8.0                                                                   \\
		GeoVirus   & News         & 08-09/2017                                           & 229                                                   & 276                                                               & 9.4                                                                   \\
		TR-News    & News         & 2009-2017                                            & 118                                                   & 324                                                               & 10.8                                                                  \\
		GeoWebNews & News         & 04/2018                                              & 200                                                   & 404                                                               & 12.6                                                                  \\
		WikToR      & Wikipedia    & 03/2016                                              & 5000                                                  & 213                                                               & 6.3                                                                   \\
		GeoCorpora & Social Media & 2014-2015                                            & 1639                                                  & 19                                                                & 2.1                                                                 \\
		Hu2014     & Web Pages    & 08/2014                                              & 134                                                   & 27                                                                & 1.3                                                                 \\
		Ju2016     & Web Pages    & 11/2016                                              & 5441                                                  & 21                                                                & 1.2                                                                 \\ \hline
	\end{tabular} \label{dataset_table}
\end{table}

%Maybe a table for datasets, with some additional attributes, such as year of the text,  number of articles, average length, toponym per article, ...} 
There also exist open and annotated historical corpora. For example,  WOTR  is a  US civil war corpus with toponyms focusing on the southern US \citep{delozier2016creating}. However, geoparsing such corpora requires special configurations such as  adding historical gazetteers and processing older languages (e.g., the texts of WOTR are from 1860s). The current version of EUPEG aims to compare geoparsers based on their default configurations and  does not include historical corpora.

%The following corpora could not be included: WoTR (DeLozier et al., 2016) due to limited coverage (southern US) and domain type (historical language, the 1860s), (De Oliveira et al., 2017) contains fewer than 180 locations, GeoText (Eisenstein et al., 2010) only allows for user geocoding, SpatialML (Mani et al., 2010) involves prohibitive costs, GeoSemCor (Buscaldi and Rosso, 2008) was annotated with WordNet senses (rather than coordinates).

%One important issue that is worth discussion is the definition of toponym. 
It is worth noting that the definition of toponym can vary across  different corpora.  It seems that  different researchers often have their own opinions on what should be considered as toponyms and what should not. This definition difference affects the ground-truth toponym annotation in a corpus. For example, \textit{LGL} considers demonyms (e.g., Canadian) as toponyms and annotates them to point coordinates (e.g., the center of Canada), whereas \textit{GeoVirus} does not annotate building names, point-of-interest (POI) names, street names, and river names.
%is that different datasets may have their own definitions of toponyms. Consequently, toponym annotations across datasets are not consistent. For example, denomyns are considered as toponyms in LGL, while river names are not considered as toponyms in GeoVirus. 
In a recent work, \cite{gritta2018pragmatic} provided a pragmatic taxonomy of toponyms which further divided toponyms into \textit{literal} and \textit{associative} toponyms with 13 sub categories.  
%However, reaching a consensus can be difficult. For example, \cite{gritta2018pragmatic} suggested that denomyns, such as \textit{Canadian}, should be annotated as toponyms.  On the other hand, it can be challenging to convince a geographer that \textit{Canadian} is a toponym that should be located to the center point of Canada. In addition, the Wikipedia article of \textit{Canadian} does not have a pair of coordinates associated either, unlike other Wikipedia articles describing geographic locations. 
In this work, we do not attempt to define toponym from one single perspective, but allow the datasets with different toponym definitions to co-exist. Such  diversity allows  users and researchers to see the different performances of a geoparser across corpora. One can then choose a geoparser that performs the best on a corpus that has a toponym definition  similar to theirs. Table \ref{toponym_definition} summarizes the different annotations of toponyms contained in each of the corpora hosted on EUPEG.
\begin{table}[h]
	\caption{Toponym annotations in different corpora.}
	\begin{tabular}{rcccccc}
		\hline
		Dataset    & \begin{tabular}[c]{@{}c@{}}Admin\\ units \\ (cities, \\ towns,\\  ...)\end{tabular} & \begin{tabular}[c]{@{}c@{}}Natural \\ features \\ (rivers, \\ mountains, \\ ...)\end{tabular} & \begin{tabular}[c]{@{}c@{}}Facilities\\ (buildings, \\ roads, \\ airports, \\ ...)\end{tabular} & \begin{tabular}[c]{@{}c@{}}Demonyms\\ (Canadian, \\ Syrian,\\ American, \\ ...)\end{tabular} & \begin{tabular}[c]{@{}c@{}}Metonymies\\ (London \\ announced\\  a new policy\\…)\end{tabular} & \begin{tabular}[c]{@{}c@{}}Modifiers \\ (Spanish \\ sausage,\\ UK beef, \\ …)\end{tabular} \\
		
		 \hline
		LGL        & \checkmark                                                                                    & \checkmark                                                                                                   & *                                                                                               & \checkmark                                                                                              & \checkmark                                                                                                & \checkmark                                                                                             \\
		GeoVirus   & \checkmark                                                                                      &                                                                                                   &                                                                                                 &                                                                                              &                                                                                                &                                                                                             \\
		TR-News    & \checkmark                                                                                      &                                                                                                   & *                                                                                               &                                                                                              &                                                                                                & \checkmark                                                                                             \\
		GeoWebNews & \checkmark                                                                                      & \checkmark                                                                                                   & \checkmark                                                                                                 & \checkmark                                                                                              & \checkmark                                                                                                & \checkmark                                                                                             \\
		WikToR     & \checkmark                                                                                      &                                                                                                   &                                                                                                 &                                                                                              &                                                                                                &                                                                                             \\
		GeoCorpora & \checkmark                                                                                      & \checkmark                                                                                                   & \checkmark                                                                                                 &                                                                                              &                                                                                                &                                                                                             \\
		Hu2014     & \checkmark                                                                                      &                                                                                                   &                                                                                                 &                                                                                              &                                                                                                &                                                                                             \\
		Ju2016     & \checkmark                                                                                      &                                                                                                   &                                                                                                 &                                                                                              &                                                                                                &                                                                                             \\ \hline
	\end{tabular} \label{toponym_definition}
\footnotesize
* indicates that the dataset contains toponym annotations in that category but does not provide geographic coordinates for the annotated toponyms.
\end{table}
\normalsize
As can be seen, all datasets include administrative units in their annotations but have different coverage on other types of entities. Some of these differences come from the corpus building process (e.g., \textit{WikToR}, \textit{Hu2014}, and \textit{Ju2016} were automatically constructed based on the names of cities and towns only), while some others originate from the different views of researchers on the definition of toponym. It seems that domain knowledge plays a major role in the annotation of toponyms. For example, \textit{GeoCorpora} is a dataset contributed by geographers, and its annotations contain only the names that can be pinned down to a certain location on the surface of the Earth. By contrast, \textit{GeoWebNews}, \textit{TR-News}, and \textit{LGL} are contributed by linguists and computer scientists who tend to annotate any terms that may have a geographic meaning (e.g., ``Canadian" and ``Spanish sausage").

%Canadian does not have a pair of coordinates in Wikipedia articles.

%\cite{leidner2006evaluation} focused on simple toponyms (named places), while \cite{mani2010spatialml} annotated places without names (e.g., this road and that building) as well as relative places (e.g., north of that building).

%Geocorpora [82] is a Twitter-based geoparsing corpus with around 6,000 toponyms with buildings and facilities annotated. The authors acknowledge that toponyms are frequently used in a metonymic manner, however, these cases have not been annotated after browsing the open dataset. 

%Present-day geography. We work with present-day earth geography (more precisely, with a February 2003 snapshot of the resources described below). Geographic names change often due to changes of administration boundaries or linguistic developments.

\subsection{Geoparsers}
We select  geoparsers for EUPEG using the following criteria. First, the selected geoparsers should function as end-to-end systems, i.e., they can take  textual documents as the input and output spatial coordinates. %Some studies only provide methods for one step of geoparsing, e.g., toponym resolution, and these methods are not included. 
Second, for academic geoparsers, the accompanying papers should be published after 2010 and they should provide publicly accessible API or downloadable software packages. Due to  technological advancements, geoparsers developed before 2010 generally do not have performances close to the state of the art, and their source codes can be hard to obtain and may not run on a modern OS. Third, for industrial geoparsers, they should provide an API that either allows free access or has a permissive rate limitation for free requests. These three criteria follow the foundational work of \cite{gritta2018s}, and 
the  geoparsers below are provided on EUPEG. 

\begin{itemize}
	\item \textit{GeoTxt}. GeoTxt is an academic geoparser developed by the GeoVISTA center of Pennsylvania State University \citep{karimzadeh2013geotxt,karimzadeh2019geotxt}. It was initially designed to geoparse microblogs (e.g., tweets), but can be applied to longer texts as well. GeoTxt provides a publicly accessible and free API at \url{http://www.geotxt.org}, and is being  maintained by its researchers. EUPEG does not host a local instance of GeoTxt but connects to  its  API. An advantage of such an online connection is that new updates of GeoTxt will be  reflected on EUPEG. On the flip side, EUPEG cannot use GeoTxt when its online service is down. EUPEG connects to version 2.0 of GeoTxt which uses its local GeoNames gazetteer deployed in July 2017.
	
	\item \textit{The Edinburgh Geoparser.} The Edinburgh Geoparser is an academic geoparser developed by the Language Technology Group (LTG) at The University of Edinburgh \citep{alex2015adapting}. A publicly available package of this geoparser is provided at: \url{https://www.ltg.ed.ac.uk/software/geoparser}. EUPEG hosts version 1.1 of the Edinburgh Geoparser which uses  the online service of GeoNames as its gazetteer. While supported on Linux and MacOS, it cannot run on Windows. 
	
	\item \textit{TopoCluster.} TopoCluster is an academic geoparser developed by \cite{delozier2015gazetteer} at the University of Texas at Austin. It performs geoparsing based on the geographic profiles of words characterized by the local Getis-Ord $G_i^*$ statistic. While their methodology focuses on toponym resolution, their source code  (\url{https://github.com/grantdelozier/TopoCluster}) provides an end-to-end system for completing both steps of geoparsing. TopoCluster does not provide an official version number. We host its latest version shared on GitHub which was updated in November 2016. 
	
	\item \textit{CLAVIN.} Cartographic Location And Vicinity INdexer is an open source geoparser that employs Apache OpenNLP Name Finder for toponym recognition, and a number of heuristics and fuzzy search for toponym resolution. CLAVIN does not come with an academic paper, but its  descriptions and source code can be obtained from GitHub (\url{https://github.com/Berico-Technologies/CLAVIN}) and Maven Central. We host CLAVIN 2.1.0 on EUPEG and it employs a local GeoNames gazetteer deployed in April 2019.
	
	\item \textit{Yahoo! PlaceSpotter.}  Yahoo! PlaceSpotter is an industrial geoparser which offers an online REST API. %It allows  developers to pass in any textual content and get back geographic entities and their coordinates. 
	As a proprietary geoparser, PlaceSpotter does not describe the exact methods behind but provides some descriptions on its functions and outputs at: \url{https://developer.yahoo.com/boss/geo/docs/PM_KeyConcepts.html}. PlaceSpotter is requested via YQL (Yahoo! Query Language), and its rate limit for free requests is relatively permissive (2,000 calls per hour; a corpus with 5,000 entries can be parsed within 3 hours).  EUPEG connects to  Yahoo! PlaceSpotter via its online REST API which employs its Where-on-Earth ID (WOEID)  for referencing places.
	
	\item \textit{CamCoder}. CamCoder is an academic geoparser developed by \cite{gritta2018melbourne} at the Language Technology Lab of the University of Cambridge. CamCoder is a deep learning based geoparser that leverages Convolutional Neural Networks (CNNs) with global maximum pooling and map-based word vector representations. The source code of CamCoder is available at: \url{https://github.com/milangritta/Geocoding-} \url{with-Map-Vector}. Running CamCoder requires   configurations on the local computing environment to include deep learning libraries, such as Tensorflow and Keras. EUPEG hosts the latest version of CamCoder shared on GitHub (updated in September 2018) which uses its local GeoNames gazetteer prepared in July 2018.
\end{itemize}

%\hl{ Based on these three criteria, the geoparsers provided on EUPEG can be further divided into two types: 1). The systems are designed specifically for geoparsing. They can be used after downloading and deployment without additional programming steps. 2). The systems are converted from a general NER tool by toponym-only limitation during the recognition step and spatial footprint linkage afterward. These conversions require extra programming steps and configurations. }

%\hl{In addition to specialized geoparsers, some general NER or entity linking tools can also be used for geoparsing.  A general NER tool could be converted to a geoparser by limiting the recognized entities to toponyms only and  assigning spatial footprints to toponyms using a gazetteer. Previous research has shown that directly assigning place names to the instances with the highest populations is a strong baseline for geoparsing and can even beat some more complex methods \mbox{\citep{speriosu2013text,delozier2015gazetteer,gritta2018pragmatic}}. The general NER tool based on an external knowledge base  can also be used as a geoparser by linking the recognized toponyms to the corresponding coordinates or other spatial information in the knowledge base. DBpedia Spotlight \mbox{\citep{mendes2011dbpedia,daiber2013improving}}  is one typical example that can recognize place names and link them to their corresponding DBpedia (the Semantic Web version of Wikipedia) pages which contain coordinates.}

In addition to the six specialized geoparsers above, EUPEG also provides three geoparsing systems extended from general NER tools (e.g., Stanford NER). These systems are included because previous research argued that geoparsing can be considered as a sub task of NER, and a geoparser can be developed by limiting the entities recognized by an NER tool to toponyms and adding spatial footprints via a gazetteer \citep{inkpen2015location}. Thus, including these NER-based geoparsing systems can help provide more comprehensive comparisons. The following three systems are hosted on EUPEG:

%The motivations of hosting these systems are: (1) geoparsing is sometimes considered as a sub task of NER; and (2) previous research has shown that assigning recognized place names to the instances with the highest populations is a strong baseline for geoparsing  \citep{speriosu2013text,delozier2015gazetteer,gritta2018pragmatic}. Thus, adding these NER-based geoparsing systems to EUPEG enables more comprehensive }
	
	%An NER tool can be converted to a geoparser by limiting the recognized entities to toponyms and  assigning spatial footprints to toponyms using a gazetteer. We host these NER-based geoparsers because previous research has shown that directly assigning place names to the instances with the highest populations is a strong baseline for geoparsing  \citep{speriosu2013text,delozier2015gazetteer,gritta2018pragmatic}. Thus, we provide the following three NER-based geoparsers on EUPEG as well

\begin{itemize}
	\item \textit{Stanford NER + Population.} Stanford NER is a powerful and open source  NER tool developed by the Stanford Natural Language Processing Group. It has been used in numerous previous studies, including some specialized geoparsers, such as GeoTxt and TopoCluster. We extend Stanford NER to a geoparsing system by  assigning the recognized toponyms to the place instances with highest populations. This simple heuristic is used because previous research has shown that assigning place names to the instances with the highest populations is a strong baseline for geoparsing and can sometimes surpass more complex models \citep{speriosu2013text,delozier2015gazetteer,gritta2018pragmatic}. We use  Stanford CoreNLP toolbox (version 3.9.2) integrated with the online service of GeoNames. %,  a comprehensive gazetteer with over eleven million place names and widely used in existing studies \citep{lieberman2010geotagging,karimzadeh2013geotxt,alex2015adapting,doi:10.1080/13658816.2017.1368523,gritta2018melbourne,kamalloo2018coherent}. 
	%The toponym recognized by Stanford NER is  then located to the place that has the highest population based on the data in GeoNames. 
	
	\item \textit{spaCy NER + Population.} spaCy is a free and open source library for natural language processing tasks in Python. Released in 2014, it has already been used in many studies and applications due to its good performance \citep{choi2015depends,jiang2016evaluating}. We integrate spaCy NER (version 2.0.18) with the online service of GeoNames, and assign the recognized toponyms to the place instances with the highest populations.
	
	\item \textit{DBpedia Spotlight.} DBpedia Spotlight is a general named entity recognition and linking (NERL) tool \citep{mendes2011dbpedia,daiber2013improving}. This type of tool not only recognizes entities from texts but also links them to the corresponding URLs in a knowledge base (such as DBpedia) which provides geographic coordinates for the recognized places. We convert DBpedia Spotlight into a geoparser by limiting the output to  toponyms and extracting their coordinates from DBpedia pages via the \textit{geo:lat} and \textit{geo:long} properties. While there exist other NERL tools such as AIDA \citep{hoffart2011robust} and TagMe \citep{ferragina2010tagme}, DBpedia Spotlight is a  widely-used NERL tool whose performance is among the state of the art \citep{van2013learning,cornolti2013framework,usbeck2014agdistis}. EUPEG hosts DBpedia Spotlight 1.0.0 with coordinates retrieved from online DBpedia pages.
\end{itemize}  

In total, EUPEG provides nine geoparsing systems for comparative experiments. Table \ref{geoparser_table} summarizes these systems and their main components.    
\begin{table}[h]
	\centering
	\caption{Geoparsing systems hosted on EUPEG and their main components.}
	\begin{tabular}{rccc}
		\hline
		Geoparser                                                           & \begin{tabular}[c]{@{}c@{}}Toponym\\ Recognition\end{tabular}               & \begin{tabular}[c]{@{}c@{}}Toponym\\ Resolution\end{tabular}           & Gazetteer                                                           \\ \hline
		\begin{tabular}[c]{@{}r@{}}   GeoTxt \\ \small{ (Version 2.0)} \end{tabular}                                                             & Stanford NER                                                                & Heuristic rules                                                        & 	\begin{tabular}[c]{@{}c@{}}  GeoNames \\   \small{(July 2017)}   \end{tabular}                                                        \\ \hline
		\begin{tabular}[c]{@{}r@{}}Edinburgh\\ \small{(Version 1.1)}\end{tabular}       & LT-TTT2                                                                     & Heuristic rules                                                        & \begin{tabular}[c]{@{}c@{}} GeoNames   \\ \small{(Online)}
		\end{tabular}                                                         \\ \hline
		\begin{tabular}[c]{@{}r@{}}TopoCluster   \\     \small{(Nov. 2016)}\end{tabular}                                                     & Stanford NER                                                                & \begin{tabular}[c]{@{}c@{}}Geo-profiles\\ of words\end{tabular}        & \begin{tabular}[c]{@{}c@{}}GeoNames+\\ Natural Earth \\ \small{(Nov. 2016)} \end{tabular}  \\ \hline
		\begin{tabular}[c]{@{}r@{}}CLAVIN  \\ \small{ (Version 2.1.0)} \end{tabular}                                                            & \begin{tabular}[c]{@{}c@{}}Apache \\ OpenNLP\end{tabular}                   & Heuristic rules                                                        & \begin{tabular}[c]{@{}r@{}}GeoNames     \\ \small{ (Apr. 2019)}   \end{tabular}                                                     \\ \hline
		\begin{tabular}[c]{@{}r@{}}Yahoo!\\ PlaceSpotter  \\ \small{ (Online)}   \end{tabular}       & Proprietary                                                                 & Proprietary                                                            & \begin{tabular}[c]{@{}c@{}}WOEID\\ (Where on Earth ID)  \\ \small{ (Online)}   \end{tabular} \\ \hline
		\begin{tabular}[c]{@{}c@{}}CamCoder   \\ \small{ (Sept. 2018)}   \end{tabular}                                                         & spaCy NER                                                                   & \begin{tabular}[c]{@{}c@{}}CNNs+Map-based\\ word vectors\end{tabular} & \begin{tabular}[c]{@{}c@{}}GeoNames  \\ \small{ (July 2018)}   \end{tabular}       \\ \hline
		\begin{tabular}[c]{@{}r@{}}Stanford NER\\ + Population \\ \small{ (Version 3.9.2)} \end{tabular} & Stanford NER                                                                & \begin{tabular}[c]{@{}c@{}}Highest \\ population\end{tabular}          & \begin{tabular}[c]{@{}c@{}}GeoNames  \\ \small{ (Online)} \end{tabular}                                                            \\ \hline
		\begin{tabular}[c]{@{}r@{}}spaCy NER\\ + Population \\ \small{ (Version 2.0.18)} \end{tabular}    & spaCy NER                                                                   & \begin{tabular}[c]{@{}c@{}}Highest \\ population\end{tabular}          & \begin{tabular}[c]{@{}c@{}}GeoNames  \\ \small{ (Online)}   \end{tabular}                                                            \\ \hline
		\begin{tabular}[c]{@{}r@{}}DBpedia\\ Spotlight \\ \small{ (Version 1.0.0)}\end{tabular}         & \begin{tabular}[c]{@{}c@{}}LingPipe Exact\\ Dictionary Chunker\end{tabular} & \begin{tabular}[c]{@{}c@{}}Context\\ similarity\end{tabular}           & \begin{tabular}[c]{@{}c@{}}DBpedia  \\ \small{ (Online)} \end{tabular}                                                               \\ \hline
	\end{tabular} \label{geoparser_table}
\end{table}  
These systems include six specialized geoparsers that can be directly deployed, and three baseline systems that are extended from general NER or NERL tools via further developments and gazetteer configurations.

%- GeoTxt (Done)
%
%- Yahoo! (Done)
%
%- Edingburgh (Done)
%
%- CLAVIN (Done)
%
%- TopoCluster (Almost Done; it takes about 10 hours to install TopoCluster on a machine; The original code has a small bug and requires the configuration of PostgreSQL and PostGIS, as well as old Python library and Python packages)
%
%
%- Stanford NER + population (Done; add the population part)
%
%- SpaCy NER + population  (to be done)
%
%- Dbpedia Spotlight (Done; coordinates)
%
%- TagMe (hold)

\subsection{Performance metrics}

EUPEG provides a number of performance metrics based on which different geoparsers can be evaluated and compared. There is no general agreement on which metrics should be used for evaluating geoparsers. As a result, we select eight metrics that were used in a variety of previous studies. In the following, we discuss these  metrics individually.

%\citep{leidner2008toponym,cheng2010you,lieberman2010geotagging,gelernter2011geo,speriosu2013text,inkpen2015location,delozier2015gazetteer,jurgens2015geolocation,santos2015using,nesi2016geographical,purves2018geographic,gritta2018s}

\begin{itemize}
	\item \textit{Precision}. %\hl{\mbox{\citep{leidner2008toponym,lieberman2010geotagging,inkpen2015location,nesi2016geographical,gritta2018s}}}. 
	Precision measures the percentage of correctly identified toponyms (true positives) among all the toponyms recognized by a geoparser. Precision was used in previous studies, such as \citep{leidner2008toponym,lieberman2010geotagging,inkpen2015location}. Precision is calculated using the following equation:
	\begin{equation}
		Precision = \frac{tp}{tp + fp}
	\end{equation}
	where $tp$ represents \textit{true positive} and $fp$ represents \textit{false positive}.
	
	\item \textit{Recall}. %\hl{\mbox{\citep{leidner2008toponym,lieberman2010geotagging,inkpen2015location,nesi2016geographical,gritta2018s}}}. 
	Recall measures the percentage of correctly identified toponyms among all the toponyms that should be identified (i.e., the  toponyms that are annotated as ground truth). Recall was used in previous studies, such as \citep{leidner2008toponym,lieberman2010geotagging,inkpen2015location}. Recall is calculated using the equation as below:
	\begin{equation}
	    Recall = \frac{tp}{tp + fn}
	\end{equation}
	where $fn$ represents \textit{false negative}.
	
	\item \textit{F-score}. %\hl{\mbox{\citep{leidner2008toponym,lieberman2010geotagging,inkpen2015location,nesi2016geographical,gritta2018s}}}. 
	F-score is the harmonic mean of precision and recall. F-score is high when both precision and recall are fairly high and is low if either of the two is low. F-score was used in previous studies, such as \citep{leidner2008toponym,lieberman2010geotagging,inkpen2015location}.  F-score is calculated using the equation below:  
	\begin{equation}
      \fscore = 2\cdot\: \frac{Precision \times Recall}{Precision + Recall}
	\end{equation}
	F-score is also called F-measure or F$_1$-score. %From a perspective of macro and micro F-scores \citep{cornolti2013framework}, the F-score implemented in EUPEG is micro F-score since it does not first compute F-scores for individual textual entries and then average the F-scores. Micro F-score is often used in geoparsing studies \citep{lieberman2012adaptive,gelernter2013algorithm,gritta2018s}.
	
	\item \textit{Accuracy}. %\hl{\mbox{\citep{leidner2008toponym,gelernter2011geo,speriosu2013text,gritta2018s}}}. 
	This metric is suitable for measuring performances on those corpora that do not have all toponyms annotated. For example, both \textit{WikToR} and \textit{Ju2016}  only annotate  a subset of all the toponyms mentioned in the text. In these situations, precision, recall, and F-score are no longer suitable, since we do not have all toponyms annotated. Accuracy can be used to quantify the percentage of the annotated toponyms that are also recognized by a geoparser. Accuracy was used in previous studies, such as \citep{gelernter2011geo,karimzadeh2016performance,gritta2018s}.  It is calculated using the equation below:
	\begin{equation}
	Accuracy = \frac{|Annotated \cap Recognized|}{ |Annotated|}
	\end{equation}
	where $Annotated$ represents the set of toponyms provided in the annotation, and $Recognized$ represents the set of toponyms recognized by the geoparser.
\end{itemize}

Precision, recall, F-score, and accuracy quantify the ability of a geoparser in correctly recognizing place names from texts rather than geo-locating these names. Accordingly, they measure the performance of a geoparser in the toponym recognition step. The establishment of matching between ground-truth annotations and geoparsing outputs is a topic that is worth discussing since it can directly affect the obtained  measures. Previous work has discussed both \textit{exact matching} and \textit{inexact matching} \citep{gritta2018s}. For a sentence such as ``The Town of Amherst has been a leader in providing online geographic information", a geoparser may recognize ``Amherst" as a toponym, while the ground-truth annotation  may be ``Town of Amherst". For \textit{exact matching}, this will be considered as both a $fp$ and a $fn$, since the output of the geoparser does not match the ground truth. For \textit{inexact matching}, it will be considered as a $tp$. We adopt \textit{inexact matching}  to accommodate such syntactically inconsistent but semantically meaningful outputs, and use the same implementation as in \citep{gritta2018s} for determining matches.  

To measure the performance of a geoparser in geo-locating toponyms, the following four metrics are provided on EUPEG. 
\begin{itemize}
	\item \textit{Mean Error Distance (MED)}. %\hl{\mbox{\citep{cheng2010you,speriosu2013text,delozier2015gazetteer,santos2015using,gritta2018s}}}. 
	MED computes the mean of the Euclidean distances between the annotated location and the location output by a geoparser. MED was used in previous studies, such as \citep{cheng2010you,speriosu2013text,santos2015using}. It is calculated using the equation below:
	\begin{equation}
		\mathit{MED} = \frac{\sum_{i=1}^{N} \sqrt{(x_i - x_i')^2 + (y_i - y_i')^2}  }{N}
	\end{equation} 
	where $N$ is the number of annotated toponyms that are recognized and geo-located by a geoparser. $(x_i, y_i)$ is the annotated coordinates, and $(x_i', y_i')$ is the geoparsed coordinates. The toponyms, which are only in the geoparsing output or only in the annotations, are not included in computing MED; those mismatches are evaluated by the previous four metrics.

	\item \textit{Median Error Distance (MdnED)}. %\hl{\mbox{\citep{speriosu2013text,delozier2015gazetteer,santos2015using,gritta2018s}}} 
	MED is sensitive to outliers which means a small number of geoparsed toponyms that are located far away from their ground-truth locations can largely distort the evaluation result. MdnED computes the median value of the error distances and is robust to outliers. MdnED was used in previous studies, such as \citep{speriosu2013text,delozier2015gazetteer,santos2015using}. MdnED is calculated as below:
    \begin{equation}
    	\mathit{MdnED} = \mathit{Median}(\{ed_i | ed_i = \sqrt{(x_i - x_i')^2 + (y_i - y_i')^2},  i \in [1, N] \})
    \end{equation}
    where $ed_i$ represents the $i$th error distance.
	
	\item \textit{Accuracy@161.} This metric calculates the percentage of the toponyms that are geo-located within 161 kilometers (100 miles) of  the ground truth locations. Accuracy@161 was used in previous studies, such as \citep{cheng2010you,delozier2015gazetteer,gritta2018s}. A main motivation of having this metric is that the geographic coordinates of a place in a gazetteer used for geoparsing may be different from the annotated coordinates. Thus, an error distance  can exist even when a geoparser correctly resolves a toponym to the right place instance. Accuracy@161 considers the result as correct as long as the resolved location is within 100 miles of the annotated location. This metric is calculated as below.
	\begin{equation}
		 \mathit{Accuracy@161} = \frac{|\{ed_i | ed_i = \sqrt{(x_i - x_i')^2 + (y_i - y_i')^2},  i \in [1, N], ed_i \leq 161 \: km \}|}{N}
	\end{equation} 
	
	\item \textit{Area Under the Curve (AUC)}. %\hl{\mbox{\citep{jurgens2015geolocation,gritta2018s}}}. 
	AUC is a metric that  quantifies the overall deviation between geoparsed locations and ground-truth annotations. AUC is computed by first plotting a curve of the normalized log  error distance and then calculating the total area under the curve. AUC was used in previous studies, such as \citep{jurgens2015geolocation,gritta2018s,gritta2018melbourne}. Figure \ref{auc_figure} shows an example of the  error distance curve.
	 \begin{figure}[h]
	 	\centering
		\includegraphics[width=0.8\textwidth]{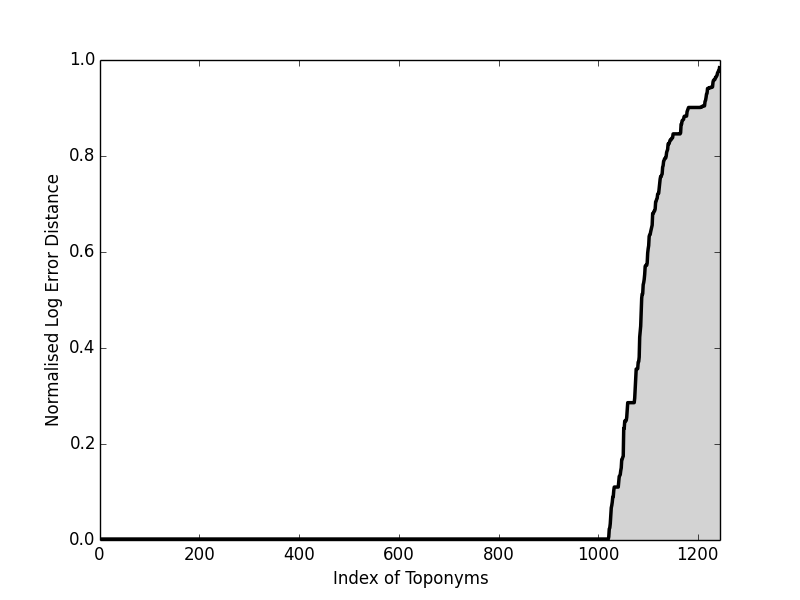}
		\caption{An illustration of AUC for quantifying the overall  error distance of a geoparser.} \label{auc_figure}
		%	\vspace*{-0.5cm}
	\end{figure}
	The horizontal axis represents the index of the toponyms ranked from small to large error distances. A majority of toponyms are typically located at the correct locations, and therefore have  errors as zero. However, once the error distance starts to appear, it can increase rapidly. The vertical axis represents the normalized log  error distance of the geoparsed toponyms. AUC is the total area under the curve calculated using Equation \ref{auc}, where $\mathit{Max\_Error}$ is the maximum possible error distance  (half of the Earth's circumference) between the ground truth and the geoparsed location. A better geoparser should have a lower AUC. 
    \begin{equation}
	\mathit{AUC} = \int_{i=1}^{N} \frac{\ln(ed_i+1)}{\ln(\mathit{Max\_Error})} di   \label{auc}
	\end{equation} 
\end{itemize}

Calculating the error distance is a fundamental step for the four metrics above. Currently, the  error distance is calculated based on point locations only. This is because all the geographic corpora we have reviewed  contain only point-based annotations, and all the discussed geoparsing systems   only output point-based locations. To some degree, this point-based annotation and geoparsing facilitate the comparison of geoparsing outputs and ground-truth annotations: it may be more difficult to reach an agreement on how to compare a geoparsing output that contains points, lines, and polygons with a ground-truth dataset that contains, e.g., points only. However, geo-locating a toponym to a single point is not ideal when the toponym refers to a large geographic area (e.g., a country or a river). Geographic scale further complicates this issue: it may be fine to locate a city name to a point if a study focuses on the country scale, but we may want a city name or even a neighborhood name to be represented as a polygon if the study is at the city scale. While existing geoparsers only output point-based locations, future geoparsers could provide  other geometries to represent the spatial footprints of the recognized toponyms. When  geoparsers and annotated datasets with various  footprints have become available, EUPEG could be extended  with error distances calculated using other methods such as Fréchet distance and Hausdorff distance.

In sum, EUPEG  provides eight  metrics for quantifying the performances of geoparsers. Four of these metrics examine the percentage of the toponyms correctly recognized from texts, while the other four metrics are based on the error distances between the geoparsed locations and their ground-truth locations. There also exist ranking-based metrics, such as Mean Reciprocal Rank (MRR) and normalized Discounted Cumulative Gain (nDCG)  \citep{purves2018geographic}. However, these metrics require a geoparser to output a ranked list of candidate places. Some end-to-end systems  only output one single result rather than a list of places. Thus, these ranking-based measures are not included.

%Seven evaluation metrics are currently provided on EUPEG. These metrics include \textit{precision}, \textit{recall}, and \textit{F-score}, which evaluate the capability of the geoparsers in identifying the correct place instance without considering the offsets of the geoparsed locations. The metrics, \textit{median} and \textit{mean}, evaluate how the locations identified by a geoparser deviate from the ground-truth locations (in kilometers). The metric, \textit{accuracy@161 km}, measures the percentages of the geoparsed locations that are within 161 km (100 miles) of the ground truth locations. The metric, \textit{Area Under the Curve} (AUC), quantifies the area under the distance error curve \citep{gritta2018s}. Figure \ref{fig3} shows the evaluation results based on the seven metrics.
%
%
%
%
%- Precision (Done)
%- Recall (Done)
%- F-score (Done)
%- Spatial offset (mean, median) (Done)
%- Accuracy@161 (Done)
%- error AUC (Done)
%- Others?

\subsection{Resource unification}

The corpora and geoparsers hosted on EUPEG are from different sources and are highly heterogeneous. %This heterogeneity can be seen in both the syntax and semantics of these resources. In terms of syntax,  
They can be annotated using different data formats. For example, LGL, GeoVirus, TR-News, GeoWebNews, and WikToR use Extensible Markup Language (XML), and organize data into hierarchical structures; Hu2014 and Ju2016 use   TXT or Comma-Separated Values (CSV), and organize data using simple line-by-line text  annotations with each line representing one data record; GeoCorpora uses the format of Tab-Separated Values (TSV) and one data record can be put into multiple lines if it contains more than one toponym. The outputs of geoparsers are also in different formats. GeoTxt and Yahoo! PlaceSpotter use JavaScript Object Notation (JSON) to format their outputs; the Edinburgh Geoparser employs XML;  TopoCluster uses its own text-based geoparsing output; and CLAVIN provides an API that allows a user to format the  output in a customized manner. %In terms of semantics, 
Even if the resources are in the same format, they can still use different vocabularies to organize similar content (which will be discussed in Section 3.6). %For example, both GeoVirus and WikToR use XML as their data format, but they use different tags (e.g., \textit{location} and \textit{toponym}) 
%to annotate the recognized place names. For geoparsers, while both GeoTxt and Yahoo! PlaceSpotter use the JSON format, they employ different labels %(i.e., \textit{feature} and \textit{place}) 
%to organize the geoparsing outputs. 

EUPEG serves as a platform for unifying these heterogeneous geoparsing resources. Building on the foundational work of \cite{gritta2018s}, we unify these resources in the following steps. First, we write a customized computer script for each geographic corpus to convert it into two parts: a collection of individual text files (with each file containing one text entry) and a single ground-truth text file (with each line containing the ground-truth annotation for one file in the collection). Such a design was used in \cite{gritta2018s}. While it seems to be rather an engineering design, we re-use it since EUPEG is built on the work of \cite{gritta2018s} and doing so can avoid reinventing the wheel. Second, we write a customized wrapper for each geoparser hosted on EUPEG. These wrappers convert the heterogeneous geoparsing outputs into the same format in which each line contains the recognized toponyms from one text file. Third, a comparison function is developed to compare the standardized geoparsing outputs with the ground-truth files, and measure the performances of the geoparsers by computing the eight  metrics. In sum, EUPEG unifies the heterogeneous resources by first converting them into the same formats and then comparing the performances of geoparsers based on the same metrics.

\subsection{Resource extension}
The resources on EUPEG can be extended with new  corpora and geoparsers. A newly created geographic corpus can be uploaded to EUPEG for testing the performances of  geoparsers. To enable the upload of any new corpus to EUPEG, we need an agreed format for organizing the text entries and ground-truth annotations in a new corpus. 
%Similar to connecting a new geoparser to EUPEG, uploading a new dataset also requires an agreed format for organizing  the texts and annotations in a geographic corpus.  
Although some toponym annotation languages, such as TRML \citep{leidner2006evaluation} and SpatialML \citep{mani2010spatialml}, have been proposed, many publicly shared corpora, such as LGL and WikToR, use their own formats, probably due to a lack of access to example datasets of TRML and SpatialML. Here, we specify the format of a new corpus to be connected to EUPEG based on LGL, GeoVirus, TR-News, GeoWebNews, and WikToR. Although these five corpora are all  in XML format, they employ different XML tags for organizing their content. For example,  GeoVirus  uses the XML tag $\left\langle article \right\rangle$  to represent each text entry, while WikToR uses the tag $\left\langle page \right\rangle$ to represent each entry (since the text entries are  Wikipedia pages). Similarly, TR-New uses the tag $\left\langle gaztag \right\rangle$ to provide location information obtained from a gazetteer, while GeoWebNews does not use the tag $\left\langle gaztag \right\rangle$ at all. %In addition, some corpora has its own tags, such as $\left\langle title \right\rangle$ for news article title information, which may not be found in other corpora.  
	Learning from these existing corpora, we build an XML format that has a small number of required core tags and offers the flexibility of including  optional  tags. Listing 1 shows this format. 

\begin{center}
	\begin{tabular}{c}
		\begin{lstlisting}[caption={The format for a new corpus to be uploaded to EUPEG.},captionpos=b]
<?xml version="1.0" encoding="utf-8"?>
<entries>
  <entry>
     <text>Paris is a city in Texas...</text>
     <toponyms>
       <toponym>
         <start>0</start>
         <end>4</end>
         <phrase>Paris</phrase>
         <place>
            <footprint>-95.5477 33.6625</footprint>
            <placename>City of Paris</placename>  #optional
            <placetype>ADM3</placetype>  #optional
        </place>
        ...  # other optional attributes
      </toponym>	
      <toponym>
         ...  # another annotated toponym
      </toponym>
      ...
    </toponyms>		
  </entry>
  <entry>
    ...  # another entry in the dataset
  </entry>
		...   
</entries>
		\end{lstlisting} 
	\end{tabular}
\end{center}

A  corpus to be uploaded to EUPEG will be organized  into one XML file using the format above. This file can contain multiple text $\left\langle entries \right\rangle$, and the $\left\langle entry \right\rangle$ tag is used to organize each individual data entry. The $\left\langle text \right\rangle$ tag contains the text to be geoparsed, which can be a news article, a tweet, a Web page, or others. The $\left\langle toponyms \right\rangle$ tag contains the  toponyms in the ground-truth annotation. For each ground-truth $\left\langle toponym \right\rangle$, it should contain the $\left\langle start \right\rangle$ position (in character index) and the $\left\langle end \right\rangle$ position of the toponym in the text. The $\left\langle phrase \right\rangle$ tag contains the name of the place mentioned in the text, which can be not only an official name but also a name abbreviation, a colloquial name, or other aliases. The $\left\langle place \right\rangle$ tag contains the annotated information of the place. The required core information for a $\left\langle place \right\rangle$ is $\left\langle footprint \right\rangle$ which is in the form of longitude and latitude. Other optional information, such as $\left\langle placename \right\rangle$ %(which provides the official name of a place) 
	and $\left\langle placetype \right\rangle$%(which indicates the type of a place based on a gazetteer),
	, can also be included.

A newly developed geoparser can  be connected to EUPEG and  compared with other hosted geoparsers. 
%(published as a Web service) to be connected and compared with other geoparsers based on the existing datasets hosted on the platform. In addition, EUPEG allows uploading new datasets for comparative experiments. These two functions are designed to help researchers and developers to spend more time on inventing new geoparsing methods and less time on (repeatedly) preparing datasets and baselines. 
%to evaluate the performance of a new geoparser without having to preparing all datasets and baselines.  Thus, EUPEG aims to help researchers to focus on developing new methods rather than preparing testing datasets or existing baselines, a process that has to be done repeatedly by different research groups.   
%To enable the direct connection of a new geoparser to EUPEG, 
%so that EUPEG can \textit{interpret} its geoparsing result. %needs to organize its parsing result in a way that can be recognized by EUPEG. %In addition, the new geoparser should be publicly accessible on the Web, so that EUPEG can connect and communicate with it on-the-fly. 
%While a number of geoparsers have been developed, their outputs are organized in different formats. 
To do so, one needs to make the new geoparser accessible via a REST API and organize the geoparsing output using an agreed format. As far as we know, there is no standard way for organizing geoparsing output. Accordingly, we specify the output format based on that of an existing geoparser, GeoTxt. As an academic geoparser, GeoTxt is available freely and publicly with a REST API and accompanied by scholarly publications \citep{karimzadeh2013geotxt,karimzadeh2019geotxt}. GeoTxt uses JSON to format its output. The original output of GeoTxt contains   elements specific to the used gazetteer, GeoNames, such as \textit{geoNameId} and \textit{featureCode}. These elements are not required in this format since a new geoparser may not necessarily employ GeoNames as its gazetteer. Similar to the  format of a new corpus,  we also classify the information elements in the geoparsing output as required core elements and optional ones. A geoparser can output only the four core elements for  simple implementation, or can  include additional and optional information for a comprehensive output. The proposed output format for a new geoparser is shown in Listing 2.

This format organizes the geoparsing output into a JSON object. It starts with a root attribute \textit{toponyms} whose value is an array of JSON objects. Each JSON object contains the information for a  toponym recognized by the geoparser. The attribute \textit{start} contains the start position (in character index) of the toponym, while the attribute \textit{end} contains the end position of the recognized toponym. The attribute \textit{phrase} represents the toponym mentioned in the  text which could be an official name or other alternative names. The attribute \textit{place} contains more detailed information about the recognized place. The  required element  is \textit{footprint}  %commonly seen in gazetteers \citep{hill2000core,goodchild2008introduction}. 
%One can also add other optional attributes per toponym, in addition to the  required  \textit{start}, \textit{end}, \textit{phrase}, and \textit{place}. 
which takes the value of a JSON array following the format of GeoJSON. For a typical point-based footprint, the JSON array contains the longitude and latitude of the place. %Such a design allows EUPEG to be extended to accommodate polyline- or polygon-shaped footprints in the future, although existing geoparsers only output point-based spatial footprints.
Other optional information, such as \textit{placename} and \textit{placetype}, can also be included.

\begin{center}
	\begin{tabular}{c}
\begin{lstlisting}[caption={The format for the output of a new geoparser to be connected to EUPEG.},captionpos=b]
{
  toponyms:
  [
	{
		start:0,
		end:4,
		phrase:"Paris",
		place:
		{
			footprint:[[-95.5477,33.6625]],
			placename:"City of Paris", # optional
			placetype:"ADM3" # optional     
		},
		...  # other optional attributes
	},
	{
		...  # another recognized toponym		
	},
	...
  ]
}
\end{lstlisting}
\end{tabular}
\end{center}

\subsection{Experiment archiving and search}
Another important function of EUPEG is archiving experiments. A database  is created to store  information about an experiment, such as \textit{Experiment ID},  \textit{Date and Time}, \textit{Datasets}, \textit{Geoparsers}, \textit{Metrics}, and \textit{Experiment Results}. An experiment ID is a 16-digit serial number that uniquely identifies an experiment. All other information is based on the  configurations specified by an user at the time of running an experiment. One can search  experiments based on their IDs and see their results.

The value of this function can be seen in two aspects. First, it facilitates the sharing of experiment results. A researcher or a geoparser user  can quickly share the  result of an experiment with colleagues by embedding the experiment ID in, e.g., an email. The colleagues who receive this experiment ID can  check it on EUPEG and see the experiment  results and configurations themselves.  Second, the independently-recorded experiment results provide further evidence for researchers to demonstrate  their work, and allow others to verify the outcome of a study more easily. Accordingly, EUPEG can help enhance the reproducibility and replicability of scientific research.

\subsection{Summary}
 We have presented the overall architecture, resources, and functions of EUPEG. In summary, EUPEG has the following features:  
\begin{itemize}
	\item \textbf{Comprehensiveness.} EUPEG provides eight annotated corpora, nine geoparsing systems, and eight performance metrics for evaluating geoparsers. The annotated corpora are in four different text genres; the geoparsing systems include both specialized geoparsers and those extended from general named entity recognizers; and the performance metrics include both information retrieval based metrics and those based on error distances.
	
	\item \textbf{Unification.}  EUPEG can be considered as a one-stop platform where corpora, geoparsers, and performance metrics are unified. EUPEG also unifies geoparser users and geoparser researchers: users can use EUPEG to select the most suitable geoparser for their own corpora, while researchers can leverage the hosted resources  to perform effective and efficient evaluation experiments.
	
	\item \textbf{Extensibility.}  EUPEG offers extensibility for the hosted geoparsing resources. A newly created corpus can be uploaded to EUPEG for testing the hosted geoparsers. A newly developed geoparser can  be connected to EUPEG and compared with other geoparsers.  We also provide the source code of EUPEG, and one can further extend EUPEG by adding new performance metrics or other features for evaluating geoparsers.
		% For researchers who would like to enrich EUPEG with additional performance measures, EUPEG offers its source code for downloading and one can implement one or multiple metrics in addition to the existing ones.}
	\item \textbf{Documentation.} EUPEG documents experiment results and configurations, and provides a search function for retrieving previous experiments. Such an archiving feature provides researchers with further evidence to demonstrate their research outcome. It also enables researchers and  users to share experiment results  more easily, e.g., by embedding  the experiment ID in an email.  
\end{itemize}

\section{Implementation and Analytical Evaluation}
\subsection{Implementation and demonstration}
\begin{figure}[h]
	\centering
	\includegraphics[width=0.95\textwidth]{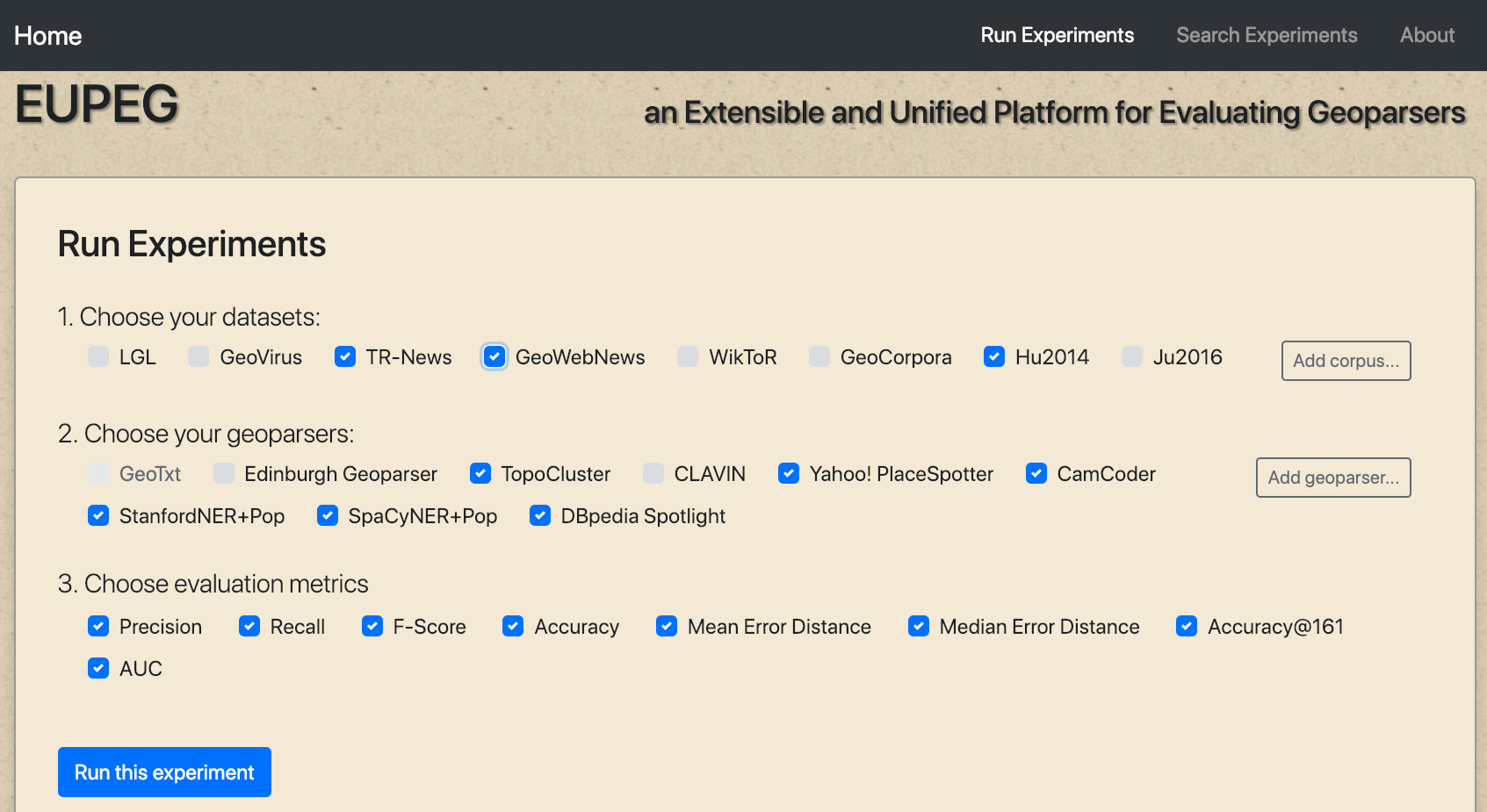}
	\caption{A screenshot of EUPEG and the (1)-(2)-(3) workflow for running an experiment.}\label{main_interface}
	%	\vspace*{-0.5cm}
\end{figure}

\begin{figure}[h]
	\centering
	\includegraphics[width=0.95\textwidth]{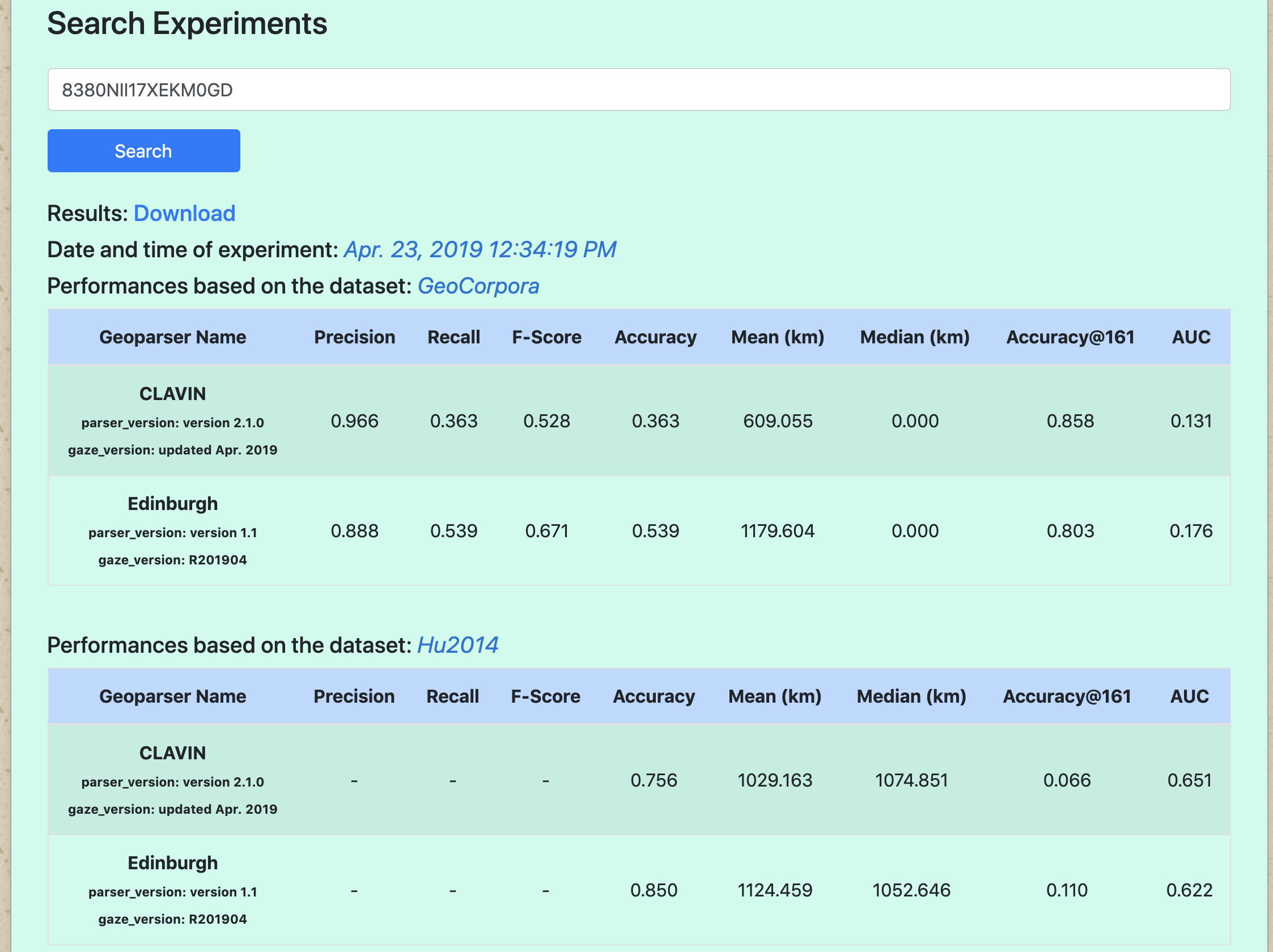}
	\caption{Searching a previous experiment and seeing its result.}\label{searchExp}
	%	\vspace*{-0.5cm}
\end{figure}

Based on the proposed architecture, we have implemented EUPEG as a Web-based platform that can be accessed online at: \url{https://geoai.geog.buffalo.edu/EUPEG}.  Figure \ref{main_interface} shows a screenshot of its main interface.
EUPEG offers a (1)-(2)-(3) workflow for conducting an experiment:  a user selects (1) datasets, (2) geoparsers, and (3) metrics, and then clicks the ``Run this experiment" button to start the experiment (Figure \ref{main_interface}). One can also click the ``Add corpus..." or ``Add  geoparser..." buttons to add their own resources. 
Once an experiment is finished, the user will be provided with an experiment ID which can then be used by the user or others to search for this experiment. Figure \ref{searchExp} shows an example of searching a previous experiment and seeing its results.  The returned results contain not only the performance information of the compared geoparsers based on the selected corpora and metrics, but also the date and time of this experiment and the versions of the geoparsers and their used gazetteers. Such information allows one to see the detailed  configuration of an experiment. A reader can also try this example by searching the experiment ID ``8380NII17XEKM0GD" on EUPEG.

EUPEG is implemented using a technology stack of multiple programming languages, software libraries, and development tools. Java JDK 11 is used on the server side  for implementing servlets, database connections, and external API requests. Javascript , HTML5, CSS3, and other libraries, such as Bootstrap and JQuery, are employed on the client side for implementing the user interface and AJAX-based  HTTP requests and responses. SQLite 3 is used for storing the experiment records, which is a light-weight, high-reliability, and public-domain database. To reduce the time of experiments and avoid running the same experiment many times, we store and re-use the results if the datasets and geoparsers selected by a user were tested in a previous experiment. Such an implementation   increases experiment efficiency and decreases  computational cost, since   running an experiment can take from hours to days depending on the selected datasets and geoparsers. In addition, we use the following approaches to keep the hosted geoparsers up-to-date. For the geoparsers that are connected to EUPEG via online APIs,  a computational thread is developed which runs in the background and automatically updates the geoparsing results of these geoparsers once per month to reflect any possible changes. For the geoparsers that are  deployed locally on our server, we plan to check their websites once every three to six months and will update our local instances  when new  stable versions have become available. While we plan to maintain EUPEG for the next few years,   resource limitation may not allow us to maintain it for a long time. Thus, we also share  the source code of EUPEG, along with the datasets under permitted licenses (e.g., GNU General Public License), on GitHub at: \url{https://github.com/geoai-lab/EUPEG}, and invite the community to further enhance and extend it.  %Researchers and end users can add new metrics or other  new features to make EUPEG better.}
	
	 %once per month and only update the result if the version changes.
%	One essential task of EUPEG is to keep the hosted geoparsers matching with the latest version. However, because of the heterogeneity among these geoparsers, purely automatic updating is not available. There are three parts of the hosted geoparsers will be subject to changing in the future. In the following, we present the strategy about how to update them.}
	%\hl{For the geoparsers that are connected to EUPEG via APIs, we update their result once per month to reflect any possible new updates of the geoparsers. }
	%\hl{For the geoparsers that are locally installed on the server, we will check their latest stable version once per month and only update the result if the version changes.} 
	%\hl{The GeoNames APIs are used as the reference during the toponyms resolution step of stanford NER and spacy NER based geoparsers. Since the GeoNames database updates itself everyday, we do not need to do extra work about it.}

\subsection{Analytical evaluation}

\begin{figure}[h]
	\centering
	\includegraphics[width=0.92\textwidth]{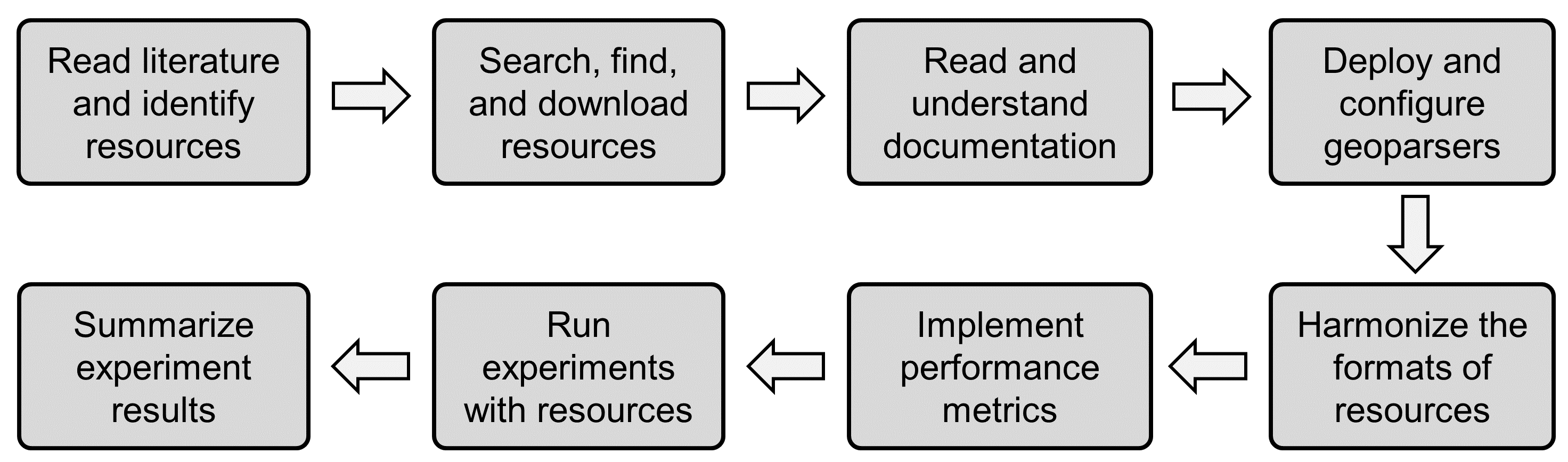}
	\caption{Typical steps for preparing datasets and geoparsers for experiments.}\label{evaluation_steps}
	%	\vspace*{-0.5cm}
\end{figure}

A main goal of EUPEG is to reduce the time that researchers have to spend in preparing datasets and baselines for experiments. This section attempts to estimate the amount of  the time that could be saved by EUPEG. One possible approach to providing such an estimate is to invite a number of  researchers, ask them to prepare all the corpora and geoparsers hosted on EUPEG by themselves, and measure the average time they spend. Such a process, however, can be very tedious for the invited researchers, and depending on their particular fields and technical skills, their used time may not represent the time that others may need for preparing these experiment resources. Here, we provide an analytical evaluation on the amount of time that could be saved  based on our own experience of developing  EUPEG and focus on the \textit{lower bound} of the time. In the following, we first analyze the steps that a research group typically has to complete if they were to prepare datasets and baselines for an experiment themselves. These identified steps are shown in Figure \ref{evaluation_steps}. We then estimate the minimum amount of time that is necessary to complete each step.% based on our own experience when working on EUPEG. 
%In the following, we discuss each step individually.  

\textit{Read literature and identify resources.} This is generally the first step, in which one studies previous research and identifies resources that can be re-used. %Here, we use the term \textit{resources} to refer to both datasets and existing geoparsers. 
For this step, EUPEG does not save much time. Although it makes various resources ready for use, researchers may still need to read the related publications to understand the  methods under the hood. EUPEG and this paper, however, can serve as an entry point for new researchers. The time that can be saved in this step is estimated as zero.

\textit{Search, find, and download resources.} After identifying  resources from the literature, one needs to obtain them. %Without EUPEG, one needs to find these resources themselves by, e.g., locating the URLs in a paper or searching online. %In some cases, one may spend quite some time searching a dataset and only find that it requires a fee (e.g., SpatialML) or is not publicly available. 
For most datasets hosted on EUPEG, we were able to obtain each of them within half an hour, thanks to the  authors who shared relevant URLs in their papers. For GeoCorpora, while the authors have kindly provided its URL, much time is still needed to rehydrate this dataset due to the data sharing restriction of Twitter.  It took us more than 5 person-hours to recover this dataset, and  additional time has to be spent in applying for a Twitter developer account before one can start to recover the dataset.  % \hlc[mybg]{In addition, we spent 8 days  in applying for a Twitter Developer Account for retrieving the tweets, and wrote multiple back-and-forth emails (including a 500-word statement on the purpose of using Twitter API) to get our application approved. Not considering the waiting time in applying for a Twitter API and the failed trials (e.g., searches that did not yield publicly accessible resources)}, 
 We estimate a minimum of 8.5 person-hours %(0.5 hour per dataset $\times$ 7 datasets + 5 hours for rehydrating GeoCorpora) for a research group to 
 for obtaining the datasets hosted on EUPEG. For the geoparsers, we were able to download the source codes or compiled versions of the Edinburgh Geoparser, TopoCluster, CLAVIN, and CamCoder within half an hour each. %, and therefore spent about 2 hours for obtaining these geoparsers. 
The other five geoparsers are either connected to EUPEG via their APIs or are further developed  based on general NER  tools. About half an hour is needed for finding each of these resources. In total, about 13 person-hours are needed for this step.

\textit{Read and understand documentation.} After the source codes of previous geoparsers are obtained, one needs to read documents and understand how to deploy and run them. %\hlc[mybg]{The code documentations have varied lengths. For example, the document of the Edinburgh Geoparser has 73 pages and 24,670 words, whereas the document of TopoCluster has 1,508 words.  A shorter document does not mean it is easier to understand the installation and use of a geoparser (it can be harder sometimes). Meanwhile, one often does not have to read the full document in order to deploy a method.} 
Background knowledge on different programming languages (e.g., Python and Java) and system architectures (e.g., REST Web services) is necessary for understanding the installation instructions. Based on our own experience, we estimate an average of two person-hours for an experienced developer to read and understand the documents of one geoparsing system. Thus, this step takes about 18 person-hours.  %(i.e., two hours per geoparser $\times$  nine geoparsing systems).

\textit{Deploy and configure geoparsers.}  %Geoparsers need to be deployed on local machines or be connected via their online APIs. 
This step  is particularly time-consuming and requires a lot of expertise. %\hlc[mybg]{ First, different geoparsers can be implemented using different programming languages or require particular operating systems.  For example, CLAVIN and Stanford NER were implemented using Java, while TopoCluster and SpaCy NER were implemented in Python. As a result, a researcher often needs to have some basic knowledge about the multiple used programming languages. For another example, Edinburgh Geoparser cannot run on Windows. While we have the expertise to deploy Edinburgh on Linux, it can take more time for a researcher who primarily works in Windows to complete such a task. Besides, one may not have the equipment, such as another computer or a virtual machine, to install the required OS. Second, there are many specific configuration requirements of particular geoparsers. For example, geoparsers available via Web APIs require one to have the expertise of handling HTTP requests and responses; a geoparser (e.g., TopoCluster) may require the installation of a database and its spatial extension, or   may  require a researcher to be familiar with certain deep learning libraries (e.g., CamCoder).  Besides these challenges, extra work is necessary if one would like to include general NER tools as baselines, such as Stanford NER, Spacy NER, and DBpedia Spotlight, where additional programming steps and gazetteer configurations are required. We estimate an average of 24 person-hours  to successfully deploy and configure one geoparser. Thus, we estimate a total of 216 person-hours for  deploying the nine geoparsers. }
First, different geoparsers can be implemented in different programming languages. %For example, CLAVIN and Stanford NER are implemented using Java, while TopoCluster and SpaCy NER are implemented in Python. 
	Accordingly, a researcher needs to have some basic knowledge on the multiple   languages in order to deploy them. Second, there exist specific configuration requirements for some geoparsers. For example, geoparsers available via Web APIs require one to have the expertise of handling HTTP requests and responses; a geoparser (e.g., TopoCluster) may require the installation of a database and its spatial extension, or   may  require a researcher to be familiar with certain deep learning libraries (e.g., CamCoder).  Third, including general NER tools as baselines requires further developments and gazetteer configurations to convert these general tools into  geoparsers. We estimate an average of 24 person-hours  to successfully deploy and configure one geoparser and thus a total of 216 person-hours.

\textit{Harmonize the formats of resources.} The annotated datasets and the outputs of geoparsers are often in different formats and structures. %For example, WikToR and GeoVirus are formatted using XML, while GeoCorpora is organized using a simple line-by-line structure. Different geoparsers can have different output formats as well,  such as JSON or customized formats. %  \hl{as we mention above.} \hlc[mybg]{For example, WikToR and GeoVirus are formatted using XML, while GeoCorpora is organized using a simple line-by-line structure. Even if datasets are constructed using the same language such as XML, they can use different XML elements for annotations. The similar situation applies to the outputs of geoparsers.} 
To conduct an experiment on these heterogeneous resources, one needs to  harmonize these datasets and geoparser outputs  by writing programs to convert them into the same format. We estimate an average of three person-hours for processing one resource (a dataset or a geoparser), and in total, this step takes about 51 person-hours.

\textit{Implement performance metrics.} Performance metrics, such as precision, recall, and AUC, need to be implemented for evaluating geoparsers. %While they can be implemented based on Equation (1) to (8), 
In addition, some programming work is necessary for comparing geoparsing outputs to   ground-truth annotations. In total, we estimate eleven person-hours for completing this step (eight hours for eight metrics plus three hours for developing the code for comparing outputs with ground-truth annotations).

\begin{figure}[h]
	\centering
	\includegraphics[width=0.8\textwidth]{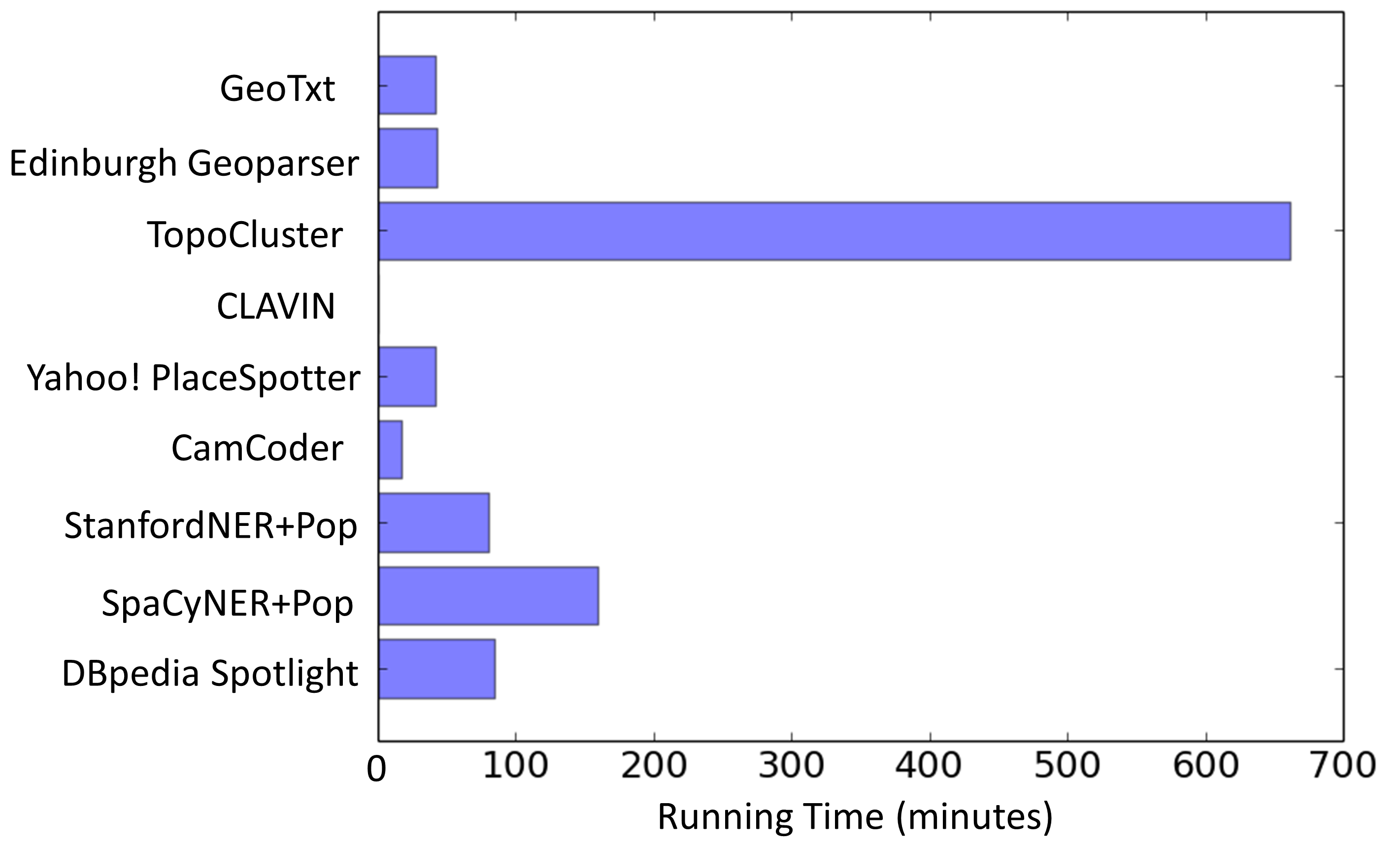}
	\caption{Running time of different geoparsers on GeoCorpora.}\label{running_time}
	%	\vspace*{-0.5cm}
\end{figure}

\begin{figure}[H]
	\centering
	\includegraphics[width=\textwidth]{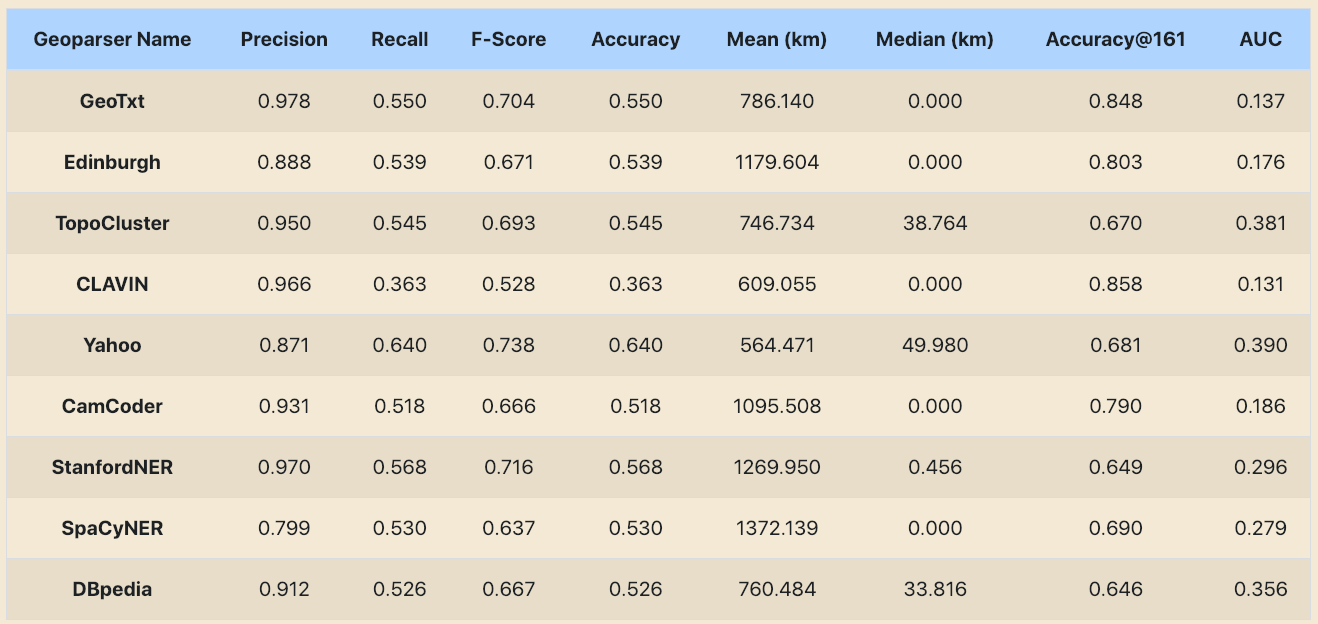}
	\caption{The performances of different geoparsers on GeoCorpora.}\label{performance_GeoCopora}
	%	\vspace*{-0.5cm}
\end{figure}

\textit{Run experiments with resources.} Once everything is  prepared , we can run experiments to obtain evaluation results. The running time of different geoparsers can vary largely. Figure \ref{running_time} reports the empirical time of the nine geoparsers on the same machine for processing the GeoCorpora dataset. TopoCluster took the longest time (660.51 minutes), while CLAVIN is the fastest geoparser that took only 0.28 minute to process the same dataset.
Longer processing time, however, does not  mean better performance. Figure \ref{performance_GeoCopora} shows the performances of different geoparsers on GeoCorpora.
The time that can be saved by EUPEG in this step is estimated as zero,  %we estimate the potentially saved running time is remarkable, since we allow a result to be reused if a geoparser has already run on a dataset. For geoparsers available via public APIs, we re-run the geoparser again once every month so that the newest update of a geoparser can be included. Despite the saved running time, 
%we consider it as zero for this step 
since one  can work on other tasks when an experiment is running.

\textit{Summarize experiment results.} When experiments are finished, one often needs to collect the obtained results and organize them into a report. For this step, we estimate the saved time as zero, since it has to be done  with or without EUPEG. 

Table \ref{save_time_table} provides a summary of the approximate number of person-hours that can be saved by EUPEG. 
\begin{table}[h]
	\caption{The estimated amount of time that can be saved by EUPEG.}
	\centering
	\begin{tabular}{lc}
		\hline
		Task for preparing experiments         & \begin{tabular}[c]{@{}c@{}}Estimated time \\ (person-hours)\end{tabular} \\ \hline
		Read literature and identify resources & 0                                                             \\
		Search, find, and download resources   & 13                                                           \\
		Read and understand documentation     & 18                                                              \\
		Deploy and configure geoparsers         & 216                                                           \\
		Harmonize the formats of resources     & 51                                                                \\
		Implement performance metrics         & 11                                                  \\
		Run experiments with resources         & 0                                                                   \\
		Summarize experiment results           & 0                                                                 \\ \hline
		Total                                  & 309                                                               \\ \hline
	\end{tabular}\label{save_time_table}
\end{table}

%\hlc[mybg]{In total, we estimate about 309 person-hours if another research group were to prepare the same resources hosted on EUPEG. Three  things need to be noted about this number. First, this 309 person-hours estimation is close to a lower bound. It assumes that the researchers have all necessary knowledge about the programming languages, OS, libraries, and databases, and does not consider the trial and errors which can cost much time. %While a research group might take longer hours to prepare these resources, we think such an estimation is  more meaningful than an overestimation on the amount of time. 
%Second, this number does not include the time of building EUPEG as a Web-based evaluation platform. The estimation of 309 person-hours assumes that these resources will only be downloaded and run locally, rather than being systematically integrated into a coherent benchmarking framework. Building EUPEG also involved conceiving the idea and designing the architecture. It took us about four months to complete the current version of EUPEG. Third, this estimation is based on the work of one single research group. If multiple groups are going to prepare these resources for evaluation experiments, EUPEG can save multiple 309 person-hours for the  research community as a whole.}

In total, we estimate  309 person-hours if another research group were to prepare the same resources hosted on EUPEG. This estimate is close to a \textit{lower bound}, as it is based on the assumption that researchers have all the necessary  knowledge and technical skills and does not take into account the time spent on trials and errors.  %While building EUPEG  took much more time than 309 person-hours, we believe it is worth to be done since it alleviates the burdens of other researchers and benefits the research community as a whole.}

\section{Conclusions and Future Work}
In this work, we present EUPEG, an Extensible and Unified Platform for Evaluating Geoparsers. With large amounts of textual data available from various sources, geoparsers have  become  increasingly important given their capabilities of extracting geographic information from textual documents.  Many studies in spatial data science and digital humanities have leveraged geoparsers  under various  contexts (e.g., disaster responses, platial studies, and event detection) to  integrates spatial and textual analysis. While a number of geoparsers were developed, they were tested on different datasets using different performance metrics. Consequently, the reported evaluation results cannot be directly compared. In addition,  a new geoparser often needs to be compared  with existing baselines to demonstrate its merits. However, preparing baselines and testing datasets can take much time and effort  from different research groups. In this context, we propose and develop EUPEG as a benchmarking platform for evaluating geoparsers and eventually enhancing spatial and textual analysis. It is implemented as a Web based and open source platform with  four major features. (1) Comprehensiveness: EUPEG provides eight open corpora, nine geoparsing systems, and eight performance metrics for evaluating geoparsers; (2) Unification: EUPEG can be considered as a one-stop platform where heterogeneous corpora, geoparsers, and  metrics are unified; (3) Extensibility: EUPEG allows the hosted resources to be extended with new corpora and geoparsers; (4) Documentation: EUPEG documents experiment results and configurations, and allows the search of previous experiments.

The main goal of EUPEG is to enable  effective and efficient comparisons of geoparsers while reducing the time that  researchers and end users have to spend in preparing datasets and baselines. Based on our analytical evaluation, EUPEG can save one single research group approximately 309 hours for preparing the same datasets,  geoparsers, and metrics. The number of saved hours will be multiplied when multiple research groups attempt to develop and compare geoparsers. While EUPEG serves as a benchmarking platform, it is not to replace project-specific evaluations necessary for highlighting certain unique features of a geoparser but to supplement existing evaluations. %Those unique features may not be demonstrated through a standard experiment performed on EUPEG. 
%Thus, the role of EUPEG should be seen as providing additional evaluation results  rather than replacing customized evaluations.   

The development of EUPEG also reveals several  issues that may need future work. First, there is a lack of commonly-agreed standards on corpus annotation. While languages, such as TRML \citep{leidner2006evaluation} and SpatialML \citep{mani2010spatialml}, were proposed, they were not adopted by the recently available and publicly shared corpora, such as LGL, WikToR, and GeoCorpora. Having a commonly-agreed  standard can facilitate the development of datasets that are more readily usable by others in future experiments. Second, a similar situation happens to the outputs of geoparsers, where a commonly-agreed  output format is not available. Most geoparsers organize their outputs in a format that they consider suitable. While it is feasible to harmonize these heterogeneous outputs by writing wrapper programs (as done in EUPEG), a standard output format can make it easier for others to use a geoparser or to combine multiple geoparsers. In this work, we have developed a simple format based on GeoTxt to allow  new geoparsers to be connected to EUPEG. However, further efforts are needed from the community to develop an agreed and standard output format for geoparsers. Third, the current version of EUPEG focuses  on English-based geoparsers and corpora only. Resources for other languages could be added in the future to support  multilingual geoparsing evaluations. With the source code shared,  new  extensions could be added to EUPEG to further enhance it and help it better serve our  community.

\section*{Acknowledgments}
The authors would like to thank Dr. Morteza Karimzadeh and Dr. Alan M. MacEachren for providing further technical information about GeoTxt. We thank the anonymous reviewers for their constructive comments and suggestions.

\section*{References}

\bibliography{reference}

\end{document}